\documentclass[12pt]{article}
\usepackage{amssymb,amsmath,bm,epsfig,comment}

\renewcommand{\thefootnote}{\fnsymbol{footnote}}
\newcommand{\im}{{\rm Im\,}}
\newcommand{\re}{{\rm Re\,}}
\begin{document}
\title{}

\title{
\begin{flushright}
\begin{minipage}{0.2\linewidth}
\normalsize
WU-HEP-15-08 \\*[50pt]
\end{minipage}
\end{flushright}
{\Large \bf 
Superfield description of $(4+2n)$-dimensional SYM theories and their mixtures 
on magnetized tori\\*[20pt] } }

\author{
Hiroyuki~Abe\footnote{
E-mail address: abe@waseda.jp}, \ 
Tomoharu~Horie\footnote{Completed the master's degree in the affiliation in 2013.} 
\ and \
Keigo~Sumita\footnote{
E-mail address: k.sumita@aoni.waseda.jp}\\*[20pt]
{\it \normalsize 
Department of Physics, Waseda University, 
Tokyo 169-8555, Japan} \\*[50pt]}

\date{
\centerline{\small \bf Abstract}
\begin{minipage}{0.9\linewidth}
\medskip 
\medskip 
\small 
We provide a systematic way of dimensional reduction for 
$(4+2n)$-dimensional $U(N)$ supersymmetric Yang-Mills (SYM) theories 
($n=0,1,2,3$) and their mixtures compactified on two-dimensional tori 
with background magnetic fluxes, which preserve a partial ${\mathcal N}=1$ 
supersymmetry out of full ${\mathcal N}=2,3$ or $4$ in the original 
SYM theories. It is formulated in an $\mathcal N=1$ superspace respecting 
the unbroken supersymmetry, and the four-dimensional effective action 
is written in terms of superfields representing $\mathcal N=1$ vector 
and chiral multiplets, those arise from the higher-dimensional SYM theories. 
We also identify the dilaton and geometric moduli dependence of matter 
K\"ahler metrics and superpotential couplings as well as of gauge kinetic 
functions in the effective action. The results would be useful for various 
phenomenological/cosmological model buildings with SYM theories or 
D-branes wrapping magnetized tori, especially, with mixture configurations 
of them with different dimensionalities from each other. 
\end{minipage}}

\begin{titlepage}
\maketitle
\thispagestyle{empty}
\clearpage
\tableofcontents
\thispagestyle{empty}
\end{titlepage}

\renewcommand{\thefootnote}{\arabic{footnote}}
\setcounter{footnote}{0}

\section{Introduction} 
\label{sec:int}
Supersymmetric Yang-Mills (SYM) theories in higher-dimensional spacetime have been 
attracting our attention from both theoretical and phenomenological points of view. 
First, they appear in low-energy limits of some superstring theories. 
The superstring theories are great candidates for a unified theory including the 
quantum gravity and have actively evolved for decades. 
Besides their beautiful theoretical features, their phenomenological aspects also 
have come to draw our attention. The higher-dimensional SYM theories accommodate 
plausible fields for such phenomenological studies and many works have been done 
on the basis of SYM theories so far 
(see Ref.~\cite{Ibanez:2012zz} for a review and references therein.). 

The SYM theories are also motivated by bottom-up approaches. 
It is known that, although the standard model (SM) is a successful theory to 
describe the nature of elementary particles discovered so far including the Higgs 
particle, there are some mysteries and unsatisfactory issues from a theoretical 
point of view in the SM, which may indicate the presence of new physics behind it. 
The basic ingredients of the higher-dimensional SYM theories relevant to their 
low-energy phenomenology are supersymmetry (SUSY) and extra dimensional space, 
which are known as promising candidates for the new physics. Therefore, it is 
sensible to study the higher-dimensional SYM theories as a particle physics model 
beyond the SM, even without mentioning superstring theories. 

From a phenomenological perspective, the matter field profile in the extra 
dimensional space is one of the principal issues to study in higher-dimensional 
theories. Especially, it has a potential for generating the observed intricate 
flavor structure of the SM without introducing hierarchical input parameters, 
due to the localized profile of fields in extra dimensions~\cite{ArkaniHamed:1999dc}. 
It was indicated that the toroidal compactification of SYM theories with magnetic 
fluxes~\cite{Bachas:1995ik,Angelantonj:2000hi,Blumenhagen:2000vk} yields product 
gauge groups, generations of chiral matter particles localized at different points 
on the tori, and potentially hierarchical Yukawa couplings among them~\cite{
Cremades:2004wa}. It is remarkable that all of such phenomenologically interesting 
features are derived as consequences of the existence of magnetic fluxes in extra 
dimensions. 

Due to such a fine prospect, a wide variety of phenomenological studies on the 
magnetized toroidal/orbifold compactifications has been done~\cite{Abe:2008fi, 
Abe:2008sx, Abe:2012ya, Abe:2012fj, Fujimoto:2013xha, Abe:2014soa, Abe:2014vza}. 
For example, in Refs.~\cite{Abe:2012fj,Abe:2014soa}, a semi-realistic model 
based on a ten-dimensional (10D) magnetized $U(8)$ SYM theory was proposed. 
This model contains all the SM gauge groups, fermion flavors, Higgs particles and 
their SUSY partners, those induced by magnetic fluxes in the extra-dimensional tori. 
Furthermore, the observed quark and lepton masses and mixing angles can be 
successfully generated by certain non-hierarchical input parameters and vacuum 
expectation values of relevant fields. 

The magnetic fluxes in the extra compact space is closely related to SUSY. 
Higher-dimensional SUSY theories intrinsically possess ${\mathcal N}=2,3$ 
and $4$ SUSY in terms of four-dimensional (4D) supercharges. 
From a phenomenological point of view, such an extended SUSY should be 
broken down to ${\mathcal N}=1$ or $0$ in order to yield a chiral spectrum 
in the 4D effective theory. It is remarkable that the magnetic fluxes in 
extra dimensions generically break the higher-dimensional SUSY~\cite{Bachas:1995ik}, 
and the number of remaining supercharges is determined by the flux configuration. 
Because ${\mathcal N}=1$ SUSY models, such as the minimal SUSY SM (MSSM), 
are phenomenologically and cosmologically attractive, it is worth studying 
higher-dimensional SYM theories compactified on tori with magnetic fluxes, 
those preserve ${\mathcal N}=1$ SUSY. 

In Ref.~\cite{Abe:2012ya}, the authors provided a systematic way of dimensional 
reduction for 10D $U(N)$ SYM theories compactified with such intended configurations 
of magnetic fluxes, and derived a 4D effective action written in terms of 
$\mathcal N=1$ superfields, where the unbroken $\mathcal N=1$ SUSY becomes manifest. 
Furthermore, the dilaton and geometric moduli dependences of matter K\"ahler 
metrics and superpotential couplings as well as of gauge kinetic functions 
were identified by upgrading the gauge coupling constant and torus parameters 
to supergravity (SUGRA) fields. Then the 4D effective SUGRA action was 
reconstructed which is described in the ${\mathcal N}=1$ superspace, and low-energy 
particle spectra including the effect of moduli-mediated SUSY breaking  were analyzed 
in Refs.~\cite{Abe:2012fj,Abe:2014soa} based on the effective SUGRA action. 

In this paper, we generalize the previous way of dimensional reduction for 
10D $U(N)$ SYM~\cite{Abe:2012ya} to those for $(4+2n)$-dimensional $U(N)$ SYM 
theories ($n=0,1,2,3$), and even for mixtures of them with different 
dimensionalities from each other. Such an extension would be quite 
meaningful because the various-dimensional SYM theories and their mixtures 
could arise as low-energy effective theories of D-brane systems in type II 
orientifold models (see Ref.~\cite{Blumenhagen:2006ci} for a review and 
references therein). Furthermore, it is expected in a bottom-up perspective that 
they are quite useful to construct more realistic models including hidden sectors 
for moduli stabilization and dynamical SUSY breaking, as well as sectors for 
yielding some non-perturbative effects to generate certain masses and couplings 
required phenomenologically and observationally in the visible and hidden sectors. 

The sections of this paper are organized as follows. 
In Sec.~\ref{sec:10d}, the superfield description of magnetized 10D SYM 
theories shown in Ref.~\cite{Abe:2012ya} is reviewed. In Sec.~\ref{sec:6d10d}, 
the simplest extension which consists of magnetized 6D and 10D SYM theories 
as well as their couplings is proposed and their 4D effective SUGRA action is shown. 
This is motivated by a D5/D9 brane system in type IIB orientifold models. 
The above mentioned semi-realistic model derived from a 10D SYM theory~\cite{Abe:2012fj,
Abe:2014soa} can be straightforwardly embedded into this system with a capacity for 
sequestered hidden sectors. Various combinations of $(4+2n)$-dimensional SYM 
theories can be treated in accordance with the procedure given in this section. 
Another example is shown in Sec.~\ref{sec:4d8d}, which consists of 4D SYM and 
magnetized 8D SYM theories accompanied by their couplings, 
motivated by a D3/D7 brane system. Sec.~\ref{sec:cd} 
is devoted to conclusions and discussions with some future prospects. 
A particular SUSY configuration for the mixture of 6D and 10D SYM theories 
is shown in Appendix~\ref{app:susy6d10d}

\section{Review of 10D magnetized SYM theory in ${\mathcal N}=1$ superspace}
\label{sec:10d} 
We give a review of the superfield description for 10D SYM theories with magnetized 
extra dimensions developed in Ref.~\cite{Abe:2012ya} based on Refs.~\cite{Marcus:1983wb, 
ArkaniHamed:2001tb}, which is the basis of extensions given in this paper. 
Most notations and conventions in this section follow those adopted in Ref.~\cite{Abe:2012ya}. 
We start from the following 10D SYM action with a 10D vector field $A_M$ and a 10D 
Majorana-Weyl spinor field $\lambda$ satisfying $\lambda^C=\lambda$ and 
$\Gamma^{10}\lambda=+\lambda$ ($\lambda^C$ is the charge conjugate to $\lambda$ and 
$\Gamma^{10}$ is the 10D chirality operator), 
\begin{equation}
S=\int d^{10}X\sqrt{-G}\frac1{g^2}{\rm Tr}\left[
-\frac14F^{MN}F_{MN}+\frac i2\bar\lambda\Gamma^MD_M\lambda\right], 
\label{eq:symaction}
\end{equation} 
where $X^M=(x^\mu,~x^m)$ is a 10D coordinate, 
and $M:0,\ldots,9$,  $\mu:0,\ldots,3$ and  $m:4,\ldots,9$. 
$F^{MN}$, $D^{M}$ and $\Gamma^M$ are the 10D field strength, the 10D covariant derivative 
and the 10D gamma matrix. The 10D gauge coupling $g$ is the sole parameter. 
We compactify it on three tori $(T^2)_i$ $(i:1,2,3)$ with $x^m\sim x^m+2$ 
and the 10D line element is then given by 
\begin{equation*}
ds^2=\eta_{\mu\nu}dx^\mu dx^\nu + c_{mn}dx^m dx^n, 
\end{equation*}
where $\eta_{\mu\nu}={\rm diag}(-, +, +, +)$ gives the 4D Minkowski spacetime and 
the 6D compact space metric $c_{mn}$ is written by a $(6\times6)$-matrix as 
\begin{equation*}
c=\begin{pmatrix}
c^{(1)}&0&0\\
0&c^{(2)}&0\\
0&0&c^{(3)}
\end{pmatrix}
\end{equation*}
using $(2\times2)$-matrix $c^{(i)}$ which represents the $i$-th torus metric. 
Its explicit form is given by 
 \begin{equation*}
c^{(i)}=\left(2\pi R_i\right)^2
\begin{pmatrix}
1&\re\tau_i\\
\re\tau_i&\left|\tau_i\right|
\end{pmatrix},
\end{equation*}
where $R_i$ and $\tau_i$ are the radius and the complex structure of $(T^2)_i$. 
In the following, instead of the real coordinates, 
we use a complex coordinate (vector) defined as 
\begin{equation*}
\begin{aligned}
z^i&\equiv\frac12(x^{2+2i}+\tau_ix^{3+2i}),\\
A_i&\equiv-\frac1{\im\tau_i}(\tau_i^*A_{2+2i}-A_{3+2i}),
\end{aligned}
\qquad\qquad
\begin{aligned}
\bar z^{\bar i}&\equiv\left(z^i\right)^*,\\
\bar A_{\bar i}&\equiv\left(A_i\right)^\dagger  .
\end{aligned}
\end{equation*}
We can then elicit the metric $h_{\bar i j}$ of this complex coordinate from 
\begin{equation*}
ds^2_{\rm 6D} = c_{mn}dx^m dx^n \equiv 2 h_{\bar i j} d\bar z^{\bar i}dz^j ,
\end{equation*}
and find 
\begin{equation*}
h_{\bar i j} = \delta_{\bar i j}2\left(2\pi R_i\right)^2 . 
\end{equation*}
The vielbein is also determined by 
$h_{\bar i j}= \delta_{\bar{ \rm i}{\rm j}}e_{\bar i}^{~\bar{\rm i}}e_j^{~\rm j}$ 
and it has the following form, 
\begin{equation*}
e_i^{~{\rm i}} = \sqrt2\left(2\pi R_i\right)\delta_i^{~{\rm i}}~. 
\end{equation*}

In this notation, the 10D vector field is decomposed into the 4D vector fields $A_\mu$ 
and the three complex fields $A_i$ $(i=1,2,3)$. 
We can also decompose the 10D Majorana-Weyl spinor field $\lambda$ 
into 4D spinors with respect to their chirality as $\lambda_{s_1s_2s_3}$, 
where $s_i=\pm$ represents its chirality on the $i$-th torus. 
A product $s_1s_2s_3$ must be $+$ to satisfy the 10D chirality condition 
$\Gamma^{10}\lambda=+\lambda$, and subsequently we can obtain four 4D Weyl spinors, 
$\lambda_{+++}$, $\lambda_{+--}$, $\lambda_{-+-}$ and $\lambda_{--+}$. 
We describe them simply as 
\begin{equation*}
\lambda_0=\lambda_{+++},\qquad
\lambda_1=\lambda_{+--},\qquad
\lambda_2=\lambda_{-+-},\qquad
\lambda_3=\lambda_{--+}~.
\end{equation*}

The decomposed bosonic and fermionic fields form the following (on-shell) 
supermultiplets of the 4D $\mathcal N=1$ SUSY which is a part of the full $\mathcal N=4$ SUSY, 
\begin{equation*}
V=\left\{v_\mu, \lambda_0\right\},\qquad
\phi_i=\left\{A_i, \lambda_i\right\}.
\end{equation*}
These are embedded into the 4D $\mathcal N=1$ vector superfield and 
the three 4D $\mathcal N=1$ chiral superfields as follows, 
\begin{eqnarray*}
V&\equiv& -\theta\sigma^\mu\bar\theta A_\mu
+i\bar\theta\bar\theta\theta\lambda_0-i\theta\theta\bar\theta\bar\lambda_0
+\frac12\theta\theta\bar\theta\bar\theta D,\\
\phi_i&\equiv& \frac1{\sqrt2}A_i+\sqrt2\theta\lambda_i+\theta\theta F_i~,
\end{eqnarray*}
where $\theta$ and $\bar\theta$ are fermionic supercoordinates 
of the $\mathcal N=1$ superspace. 

The 10D SYM action (\ref{eq:symaction}) 
can be rewritten with the 
superfields $V$ and $\phi$ in the $\mathcal N=1$ superspace as~\cite{Marcus:1983wb,ArkaniHamed:2001tb} 
\begin{eqnarray}
S=\int d^{10}X\sqrt{-G}\left[
\int d^4\theta {\mathcal K}+\left\{
\int d^2\theta\left(\frac1{4g^2}\mathcal W^\alpha\mathcal W_\alpha +\mathcal W\right)
+ {\rm h.c.}
\right\}
\right] ,\label{eq:symsusy}
\end{eqnarray}
where the functions $\mathcal K$, $\mathcal W$ and $\mathcal W^\alpha$ are given by 
\begin{eqnarray*}
\mathcal K &=& \frac2{g^2} h^{\bar ij}{\rm Tr} \left[
\left(\sqrt2\bar\partial_{\bar i} +\bar\phi_{\bar i}\right)e^{-V}
\left(-\sqrt2\partial_j +\phi_j\right)e^{V} +\bar\partial_{\bar i}e^{-V}\partial_je^V\right]
+\mathcal K_{\rm WZW},\\
\mathcal W &=& \frac1{g^2}\epsilon^{\rm ijk}e_{\rm i}^{~i}e_{\rm j}^{~j}e_{\rm k}^{~k} {\rm Tr} 
\left[\sqrt2\phi_i\left(\partial_j\phi_k-\frac1{3\sqrt2}[\phi_j, \phi_k]\right)\right],\\
\mathcal W_\alpha &=& -\frac14\bar D\bar D e^{-V}D_\alpha e^V~.
\end{eqnarray*}
$\partial_i$ represents the derivative with respect to $z_i$, and $D_\alpha$ and 
$\bar D_{\dot\alpha}$ 
are the supercovariant derivative and its conjugate. 
$\mathcal K_{\rm WZW}$ is the Wess-Zumino-Witten term which vanishes in the Wess-Zumino 
gauge fixing. $\epsilon^{\rm ijk}$ is the anti-symmetric tensor. 
This action remain invariant under the full $\mathcal N=4$ SUSY 
and the superspace formulation make the $\mathcal N=1$ SUSY manifest. 

This superspace formulation contains some auxiliary fields. 
Field equations for them are given by 
\begin{eqnarray}
D &=& -h^{\bar ij}\left(\bar\partial_{\bar i}A_j+\partial_j\bar A_{\bar i} 
+\frac12\left[\bar A_{\bar i}, A_j\right]\right),\label{eq:dvev}\\
\bar F_{\bar i} &=& -h_{j\bar i} \epsilon^{\rm jkl} e_{\rm j}^{~j}e_{\rm k}^{~k}e_{\rm l}^{~l}
\left(\partial_kA_l-\frac14\left[A_k, A_l\right]\right). \label{eq:fvev}
\end{eqnarray}
The $\mathcal N=1$ SUSY is preserved as long as 
vacuum expectation values (VEVs) of these auxiliary fields $D$ and $F_i$ are vanishing.

In the following, we consider a SUSY vacuum where 
the decomposed 10D fields develop their VEV as 
\begin{equation*}
\langle A_i\rangle\neq0,\qquad 
\langle A_\mu\rangle=\langle \lambda_0\rangle=\langle \lambda_i\rangle=0. 
\end{equation*}
The vanishing VEVs are required for the 4D Lorentz invariance and 
the nonvanishing one of $A_i$ is expected to satisfy $\langle D\rangle=\langle F_i\rangle=0$ with 
Eqs.~(\ref{eq:dvev}) and (\ref{eq:fvev}). 
We expand the 10D SYM action around this vacuum in the superspace formulation, that is, 
we redefine fluctuations of the fields as 
\begin{equation*}
V\rightarrow\langle V\rangle+ V,\qquad\phi_i\rightarrow\langle\phi_i\rangle+\phi_i,
\end{equation*}
where $\langle V\rangle=0$ and $\langle\phi_i\rangle=\langle A_i\rangle/\sqrt2$. 
From now on, $V$ and $\phi_i$ represent fluctuations around a nontrivial magnetized vacuum. 
We use these in the SYM action (\ref{eq:symsusy}) and expand it in powers of $V$. 
The functions $\mathcal K$ and $\mathcal W$ are then given by 
\begin{eqnarray}
\mathcal K &=& \frac2{g^2} h^{\bar ij}{\rm Tr} \left[
\bar\phi_{\bar i}\phi_j+\sqrt2\left\{\left(\bar\partial_{\bar i}\phi_j+\frac1{\sqrt2}
[\langle\bar\phi_{\bar i}\rangle,\,\phi_j] + {\rm h.c.} \right)
+\frac1{\sqrt2}[\bar\phi_{\bar i},\,\phi_j]\right\}V \right.\nonumber\\
&& \left.
+\left(\bar\partial_{\bar i}V\right)\left(\partial_jV\right) 
+\frac12\left(\bar\phi_{\bar i}\phi_j+\phi_j\bar\phi_{\bar i}\right)V^2
-\bar\phi_{\bar i}V\phi_jV\right] +\mathcal K^{(\rm D)}+\mathcal K^{(\rm br)},\nonumber\\
\mathcal W &=& \frac1{g^2}\epsilon^{\rm ijk}e_{\rm i}^{~i}e_{\rm j}^{~j}e_{\rm k}^{~k} {\rm Tr} 
\left[\sqrt2\left(\partial_i\phi_j-\frac1{\sqrt2}[\langle\phi_i\rangle, \phi_j]\right)\phi_k
-\frac23\phi_i\phi_j\phi_k\right] +\mathcal W^{(\rm F)},\label{eq:symmag}
\end{eqnarray}
where the expansion terminates at $V^2$ because the supercoordinates 
$\theta$ and  $\bar\theta$ are anticommuting two-component Weyl spinors. 
$\mathcal K^{(\rm D)}$ and $\mathcal W^{(\rm F)}$ are vanishing 
when the $\mathcal N=1$ SUSY is preserved. 
$\mathcal K^{(\rm br)}$ represents a mass term of V corresponding to 
partial gauge symmetry breaking due to the magnetic fluxes (we will explain later) 
and also contains other interaction terms. 
$\mathcal W^\alpha$ is not changed because it contains only $V$ and its VEV is vanishing. 

\subsection{Zero-mode equations}
\label{sec:zeromo}
In the toroidal compactification, 
the superfields $V$ and $\phi_i$ can be decomposed with Kaluza-Klein (KK) towers as 
\begin{eqnarray}
V(x^\mu, z^j, \bar z^{\bar j}) &=& \sum_{\bm n}\left(
f_0^{(1),n_1}(z^1,\bar z^{\bar 1})\times
f_0^{(2),n_2}(z^2,\bar z^{\bar 2})\times
f_0^{(3),n_3}(z^3,\bar z^{\bar 3})\right)\times 
V^{\bm n} (x^\mu)\nonumber\\
\phi_i(x^\mu, z^j, \bar z^{\bar j}) &=& \sum_{\bm n}\left(
f_i^{(1),n_1}(z^1,\bar z^{\bar 1})\times
f_i^{(2),n_2}(z^2,\bar z^{\bar 2})\times
f_i^{(3),n_3}(z^3,\bar z^{\bar 3})\right)\times 
\phi_i^{\bm n} (x^\mu), \label{eq:kkmodes}
\end{eqnarray}
where $\bm n = \left(n_1, n_2, n_3\right)$. 
$V^{\bm n}$ and $\phi_i^{\bm n}$ are $\bm n$-th KK modes 
and their internal wavefunctions on 
the $j$-th torus are described by $f_0^{(j)}$ and $f_i^{(j)}$, respectively. 
They have the Yang-Mills indices but we omit them here. 
The internal wavefunction is common to scalar and spinor fields included in a superfield 
as long as the SUSY is preserved, and 
their dependence on the supercoordinate appears only in $V^{\bm n}$ and $\phi_i^{\bm n}$. 

In the following, we focus on zero-modes with $n_1=n_2=n_3=0$ and 
denote their internal wavefunctions simply by $f_0^{(j)}$ and $f_i^{(j)}$ omitting $n_j=0$ 
for $j=1,2,3$, that is, $f_0^{(j)}\equiv f_0^{(j),n_j=0}$ and $f_i^{(j)}\equiv f_i^{(j),n_j=0}$. 
In the superspace action (\ref{eq:symmag}) given on a nontrivial background, 
the following zero-mode equations can be found, 
\begin{eqnarray*}
\bar\partial_{\bar i} f_0^{(i)} +\frac12[\langle\bar\phi_{\bar i}\rangle,\,f_0^{(i)}] &=&0,\\
\bar\partial_{\bar i} f_j^{(i)} +\frac12[\langle\bar\phi_{\bar i}\rangle,\,f_j^{(i)}] &=&0
\qquad{\rm for}\quad i=j,\\
\partial_{\bar i} f_j^{(i)} -\frac12[\langle\phi_i\rangle,\,f_j^{(i)}] &=&0
\qquad{\rm for}\quad i\neq j.
\end{eqnarray*}

We introduce (Abelian) magnetic fluxes and continuous Wilson lines in the extra compact space. 
The vacuum configuration $\langle\phi_i\rangle=\langle A_i\rangle/\sqrt2$ is then given by 
\begin{equation}
\langle A_i\rangle = \frac\pi{\im \tau_i} \left(M^{(i)}\bar z_{\bar i} +\bar\zeta^{(i)}\right), 
\label{eq:aconfig}
\end{equation}
where magnetic fluxes $M^{(i)}$ and Wilson lines $\zeta^{(i)}$ are $(N\times N)$-diagonal matrices 
corresponding to the $U(N)$ gauge symmetry of the SYM theory. 
Note that, each entries of $M^{(i)}$ must be integer because of 
the Dirac's quantization condition. We also expect them to satisfy the SUSY condition 
$\langle D\rangle=\langle F\rangle=0$ with Eqs.~(\ref{eq:dvev}) and (\ref{eq:fvev}). 
The Abelian $(1,1)$-form flux (\ref{eq:aconfig}) always satisfies $\langle F\rangle=0$ 
but the other $\langle D\rangle=0$ requires each entry of $M^{(i)}$ to satisfy 
\begin{equation*}
\sum_i\frac1{\mathcal A^{(i)}}m_k^{(i)} =0, 
\end{equation*}
where $m_k^{(i)}$ is the $k$-th entry of the diagonal matrix $M^{(i)}$, and 
$\mathcal A^{(i)}$ represents the area of the $i$-th torus. 

The magnetic fluxes and the Wilson-lines can break the gauge symmetry of SYM theories. 
For example, when all the $N$ entries of diagonal matrix $M^{(i)}$ take different values 
from each other, an original $U(N)$ gauge symmetry is broken down to 
a product of $N$ $U(1)$ symmetries. 
In another case when some of them take the same values, that is, 
the magnetic fluxes are given as 
\begin{equation*}
M^{(i)} = {\rm diag}(\overbrace{m_1^{(i)},m_2^{(i)},\cdots,m_{N_1}^{(i)}}^{=M_{N_1}^{(i)}},
\overbrace{m_{N_1+1}^{(i)},\cdots,m_{N_1+N_2}^{(i)}}^{=M_{N_2}^{(i)}},\cdots,
\overbrace{m_{N_1+\cdots+N_{n-1}+1}^{(i)},\cdots,m_{N_1+\cdots+N_n}^{(i)}}^{=M_{N_n}^{(i)}}),
\end{equation*}
they break the gauge symmetry as 
$U(N)\rightarrow\prod_a U(N_a)$ (Note that $M_{N_a}^{(i)}\neq M_{N_b}^{(i)}$). 
This discussion also apply to the Wilson lines. 
We use indices $a,b$ and $c$ to label unbroken gauge subgroups of $U(N)$.

We denote a bifundamental representation $(N_a,~\bar N_b)$ 
of the zero-mode $f_j^{(i)}$ by $(f_j^{(i)})_{ab}$. 
The zero-mode equations for the representation $(f_j^{(i)})_{ab}$ on the torus $(T^2)_i$ 
are given by 
\begin{eqnarray}
\left[\bar\partial_{\bar i} +\frac{\pi}{2\im\tau_i}
\left(M_{ab}^{(i)}z_i+\zeta_{ab}^{(i)}\right)\right](f_j^{(i)})_{ab} &=& 0 \qquad{\rm for}\quad i=j,
\label{eq:zeroii}\\
\left[\partial_i -\frac{\pi}{2\im\tau_i}
\left(M_{ab}^{(i)}\bar z_{\bar i}+\bar\zeta_{ab}^{(i)}\right)\right]
(f_j^{(i)})_{ab} &=& 0 \qquad{\rm for}\quad i\neq j, \label{eq:zeroij}
\end{eqnarray}
 where 
\begin{equation*}
M_{ab}^{(i)}\equiv M_{N_a}^{(i)}-M_{N_b}^{(i)},\qquad
\zeta_{ab}^{(i)}\equiv \zeta_{N_a}^{(i)}-\zeta_{N_b}^{(i)}. 
\end{equation*}

A normalizable solution of Eq.~(\ref{eq:zeroii}) is found~\cite{Cremades:2004wa} as 
\begin{equation*}
(f^{(i)}_j)_{ab} = f^{I_{ab}^{(i)}}\equiv
\left\{ \begin{array}{ll}
\displaystyle 
\Theta^{I^{(i)}_{ab}, M^{(i)}_{ab}}(\tilde z_i)
& \quad (M^{(i)}_{ab} > 0) \\
\displaystyle ({\cal A}^{(i)})^{-1/2} & 
\quad (M^{(i)}_{ab} = 0) \\
\displaystyle 
0 & \quad (M^{(i)}_{ab} < 0) 
\end{array} \right., 
\end{equation*}
where $\tilde z_i\equiv z_i+\frac{\zeta^{(i)}_{ab}}{M_{ab}^{(i)}}$ and 
\begin{equation*}
I^{(i)}_{ab}\equiv
\left\{ \begin{array}{ll}
\displaystyle 
1,\ldots,|M^{(i)}_{ab}| & \quad (M^{(i)}_{ab} > 0) \\
\displaystyle 0 & \quad (M^{(i)}_{ab} = 0) \\
\displaystyle {\rm no~ solution} & \quad (M^{(i)}_{ab} < 0) 
\end{array} \right. .  
\end{equation*}
When $M^{(i)}_{ab} > 0$, $M^{(i)}_{ab}$ normalizable zero-modes appear and 
they are labeled by the index $I^{(i)}_{ab}$. 
On the other hand, zero-modes are projected out 
by the magnetic fluxes when $M^{(i)}_{ab} < 0$. 
A vanishing magnetic flux $M^{(i)}_{ab} = 0$ induces a trivial zero-mode 
with a flat profile of wavefunction. 
The zero-mode wavefunction $\Theta^{I^{(i)}_{ab}, M^{(i)}_{ab}}$ in the above expression 
is defined by 
\begin{equation} 
\Theta^{I,M}\left(z \right)=\mathcal N_M e^{\pi i M z\im z/\im\tau}
\vartheta\begin{bmatrix}
I/M\\0\end{bmatrix}\left(Mz, M\tau\right), \label{eq:waveform}
\end{equation}
where the Jacobi-theta function is given by 
\begin{equation*}
\vartheta\begin{bmatrix}
a\\b\end{bmatrix}\left(\nu, \tau\right)=\sum_{ l\in \mathbb Z} 
e^{\pi i\left(a+ l\right)^2\tau}
e^{2\pi i \left(a+ l \right)\left(\nu +b\right)}.
\end{equation*}
The normalizations are determined by 
\begin{equation}
\int  dz_id\bar z_{\bar i} \sqrt{{\rm det}\, c^{(i)}} f^I \left(f^J\right)^* =\delta_{IJ}, 
\label{eq:normal}
\end{equation}
and it leads to  
\begin{equation*}
\mathcal N_M  = \left(\frac{2\im\tau_i|M|}{(\mathcal A^{(i)})^2}\right)^{1/4}. 
\end{equation*}

We can also describe a normalizable solution of Eq.~(\ref{eq:zeroij}) as 
\begin{equation*}
(f^{(i)}_j)_{ab} = f^{I_{ab}^{(i)}}\equiv
\left\{ \begin{array}{ll}
\displaystyle 
0 & \quad (M^{(i)}_{ab} > 0) \\
\displaystyle ({\cal A}^{(i)})^{-1/2} & 
\quad (M^{(i)}_{ab} = 0) \\
\displaystyle 
(\Theta^{I^{(i)}_{ab}, M^{(i)}_{ab}}(\tilde z_i))^* & \quad (M^{(i)}_{ab} < 0) 
\end{array} \right., 
\end{equation*}
and $|M^{(i)}_{ab}|$ normalizable zero-modes are obtained when $M^{(i)}_{ab} < 0$ 
for $i\neq j$. 

\subsection{4D effective action} 
We give a 4D effective action derived from the 10D magnetized SYM theory 
in the superspace formulation, concentrating on zero-modes of 
gauge fields of unbroken gauge subgroups $(V^{\bm n= \bm 0})_{aa}$ 
and bifundamental matter fields $(\phi_i^{\bm n= \bm 0})_{ab}$ ($a\neq b$) 
in the assumption of gauge symmetry breaking $U(N)\rightarrow\prod_a U(N_a)$ 
due to the magnetic fluxes\footnote{
We remark on the other elements, $(V^{\bm n= \bm 0})_{ab}$ $(a\neq b)$ 
and $(\phi_i^{\bm n= \bm 0})_{aa}$. 
A bifundamental representation of the gauge multiplets $(V^{\bm n= \bm 0})_{ab}$ $(a\neq b)$ 
gets  its mass corresponding to the partial gauge symmetry breaking, 
which mass should be large comparable to the compactification scale. 
The other $(\phi_i^{\bm n= \bm 0})_{aa}$ remains massless and 
we need a prescription to make them heavy or eliminate them. 
Toroidal orbifolds, for example, can eliminate these extra zero-mode~\cite{Abe:2008fi,Abe:2012fj}. }. 
In the following, we consider a case 
with $M_{ab}^{(i)}>0$ and $M_{ab}^{(j)}<0$ for $\forall j\neq i$. 
The total number of zero-modes $(\phi_i^{\bm n= \bm 0})_{ab}$ which appear 
in the 4D effective field theory is then given by 
\begin{equation*}
N_{ab} = |\prod_{i=1}^3M_{ab}^{(i)}|, 
\end{equation*}
while $(V^{\bm n= \bm 0})_{aa}$ does not feel magnetic fluxes and 
a single zero-mode with a flat wavefunction is obtained. 
We denote them simply by 
\begin{equation*}
(V^{\bm n= \bm 0})_{aa} \equiv V^a, \qquad 
(\phi_i^{\bm n= \bm 0})_{ab} \equiv g\phi_i^{\mathcal I_{ab}}, 
\end{equation*} 
where $\mathcal I_{ab} =(I_{ab}^{(1)}, I_{ab}^{(2)}, I_{ab}^{(3)})$ labels 
$N_{ab}$ zero-modes, that is, $\mathcal I_{ab} =1,2,\ldots, N_{ab}$. 
We normalize the chiral superfields $\phi_i$ by the gauge coupling constant $g$ 
for the later convenience. 

In the 4D effective field theory with these zero-modes, 
we can compute Yukawa and higher-order couplings 
as integrals of wavefunctions of the form (\ref{eq:waveform}), 
which can be performed analytically~\cite{Cremades:2004wa,Abe:2009dr}. 
We substitute the KK-mode expansion (\ref{eq:kkmodes}) in Eq. (\ref{eq:symmag}) 
and extract a part involving the zero-modes $V^a$ and $\phi_j^{\mathcal I_{ab}}$. 
That is described by 
\begin{eqnarray}
S=\int d^4 x \left[
\int d^4\theta {\mathcal K_{\rm eff}}+\left\{
\int d^2\theta\left(\frac1{4g_a^2}\mathcal W^{a,\alpha}\mathcal W^a_\alpha 
+\mathcal W_{\rm eff}\right)+ {\rm h.c.}
\right\}\right] ,\label{eq:obtain}
\end{eqnarray}
where the functions $\mathcal K_{\rm eff}$, $\mathcal W_{\rm eff}$ and 
$\mathcal W^a_\alpha$ have the following form, 
\begin{eqnarray*}
\mathcal K_{\rm eff} &=& \sum_{i,j}\sum_{a,b}\sum_{\mathcal I_{ab}} 
\tilde Z_{\mathcal I_{ab}}^{\bar i j} {\rm Tr} \left[ 
\bar\phi_{\bar i}^{\mathcal I_{ab}} e^{-V^a} \phi_j^{\mathcal I_{ab}} e^{V^a} \right], \\
\mathcal W_{\rm eff} &=& \sum_{i,j,k}\sum_{a,b,c}
\sum_{\mathcal I_{ab},\mathcal I_{bc},\mathcal I_{ca}}
\tilde\lambda^{ijk}_{\mathcal I_{ab}\mathcal I_{bc}\mathcal I_{ca}}{\rm Tr} 
\left[\phi_i^{\mathcal I_{ab}}\phi_j^{\mathcal I_{bc}}\phi_k^{\mathcal I_{ca}}\right],\\
\mathcal W_\alpha &=& -\frac14\bar D\bar D e^{-V^a}D_\alpha e^{V^a}.\qquad 
g_a = g \left(\prod_i\mathcal A^{(i)}\right)^{-1/2}~.
\end{eqnarray*}
In this expression, K\"ahler metric $\tilde Z_{\mathcal I_{ab}}^{\bar i j}$ 
and holomorphic Yukawa coupling 
$\tilde\lambda^{ijk}_{\mathcal I_{ab}\mathcal I_{bc}\mathcal I_{ca}}$ 
are determined by integrals in the 6D extra compact space 
and they can be written as 
\begin{eqnarray}
\tilde Z_{\mathcal I_{ab}}^{\bar i j} &=& 2h^{\bar ij}\label{eq:tildez}\\
\tilde\lambda^{ijk}_{\mathcal I_{ab}\mathcal I_{bc}\mathcal I_{ca}} &=& 
-\frac{2g}3\epsilon^{\rm ijk}e_{\rm i}^{~i}e_{\rm j}^{~j}e_{\rm k}^{~k} \prod_{r=1}^3 
\tilde\lambda^{(r)}_{I^{(r)}_{ab} I^{(r)}_{bc} I^{(r)}_{ca}}, \label{eq:tildelambda}
\end{eqnarray}
where 
\begin{equation}
\tilde\lambda^{(r)}_{^{(r)}I_{ab} I^{(r)}_{bc} I^{(r)}_{ca}} = 
\int dz^rd\bar z^{\bar r} \sqrt{{\rm det}\, c^{(r)}} 
f^{I_{ab}^{(r)}} f^{I_{bc}^{(r)}} f^{I_{ca}^{(r)}} . \label{eq:yukawaint}
\end{equation}
We have performed the integral in the K\"ahler metric by using Eq.~(\ref{eq:normal}). 
The calculation of Yukawa couplings (\ref{eq:yukawaint}) can also be carried out analytically 
and we summarize the results as follows, 
\begin{equation}
\tilde\lambda^{(r)}_{I^{(r)}_{ab} I^{(r)}_{bc} I^{(r)}_{ca}} = 
\left\{ \begin{array}{ll}
\displaystyle 
\tilde\lambda^{(r)}_{ab,c} & \quad (M^{(i)}_{ab} > 0) \\
\displaystyle 
\tilde\lambda^{(r)}_{bc,a} & \quad (M^{(i)}_{bc} > 0) \\
\displaystyle 
\tilde\lambda^{(r)}_{ca,b} & \quad (M^{(i)}_{ca} > 0) 
\end{array} \right.,\label{eq:3ways} 
\end{equation}
where
\begin{eqnarray}
\tilde\lambda^{(r)}_{ab,c} &=& 
{\mathcal N}_{M^{(r)}_{ab}}^{-1}{\mathcal N}_{M^{(r)}_{bc}}{\mathcal N}_{M^{(r)}_{ca}}
\sum_{m=1}^{M^{(r)}_{ab}} 
\delta_{I^{(r)}_{bc}+I^{(r)}_{ca}-m M^{(r)}_{bc},~I^{(r)}_{ab}}\nonumber\\
 && \times 
\exp \left[ \frac{\pi i}{\im\tau_r} 
\left( 
  \frac{\bar\zeta^{(r)}_{ab}}{M^{(r)}_{ab}}\im\zeta^{(r)}_{ab}
+\frac{\bar\zeta^{(r)}_{bc}}{M^{(r)}_{bc}}\im\zeta^{(r)}_{bc}
+\frac{\bar\zeta^{(r)}_{ca}}{M^{(r)}_{ca}}\im\zeta^{(r)}_{ca}
\right) \right] \nonumber\\
 && \times \vartheta 
\begin{bmatrix}
\frac{
M^{(r)}_{bc} I^{(r)}_{ca} - M^{(r)}_{ca} I^{(r)}_{bc} + m M^{(r)}_{bc} M^{(r)}_{ca}}
{M^{(r)}_{ab} M^{(r)}_{bc} M^{(r)}_{ca}} 
\\ 0 \end{bmatrix} 
\left( 
 \bar\zeta^{(r)}_{ca}M^{(r)}_{bc} - \bar\zeta^{(r)}_{bc}M^{(r)}_{ca}, 
-\bar\tau_r M^{(r)}_{ab}M^{(r)}_{bc}M^{(r)}_{ca} \right)~.\label{eq:yukawaxxx} 
\end{eqnarray}
This expression is obtained in the case with 
$M^{(r)}_{ab} M^{(r)}_{bc} M^{(r)}_{ca} >0$. 
In another case with vanishing magnetic fluxes, that is, 
$M^{(r)}_{ab} M^{(r)}_{bc} M^{(r)}_{ca} = 0$, 
the integral in Eq.~(\ref{eq:yukawaint}) induces a simple factor, 
$\tilde\lambda^{(r)}_{I^{(r)}_{ab} I^{(r)}_{bc} I^{(r)}_{ca}}=(\mathcal A^{(r)})^{-1/2}$. 

\subsection{Effective supergravity and moduli multiplets} 
We have obtained the 4D effective action based on the 10D SYM theories 
in the magnetized toroidal compactification. 
We can read the action in the framework of supergravity (SUGRA) 
introducing the moduli fields. 
The 10D SYM theory is described with a global SUSY but 
its 4D effective action has remnants of local structure of the SUSY, 
such as, the 10D gauge coupling $g$ and the torus parameters $R^{(i)}$ and $\tau_i$. 
The moduli fields are related to complex and K\"ahler structures,  
and the 10D dilaton $\phi_{10}$ determines the gauge coupling 
as $g=e^{\langle\phi_{10}\rangle/2}$. 
Thus, we can define the moduli and dilaton superfields by the remnant parameters 
in the toroidal compactification as follows, 
\begin{equation}
\re \langle S\rangle = e^{-\langle\phi_{10}\rangle}\prod_{i=1}^3\mathcal A^{(i)}, \qquad
\re \langle T_i\rangle = e^{-\langle\phi_{10}\rangle}\mathcal A^{(i)}, \qquad
\langle U_i\rangle = i\bar\tau_i~.\label{eq:moduli} 
\end{equation}

The obtained 4D effective action should fit into the following general form of the action 
for 4D $\mathcal N=1$ conformal SUGRA with the moduli superfields,  
\begin{eqnarray}
S &=& \int d^4x\sqrt{-g^C}\left[-3\int d^4\theta\bar CCe^{-K/3}\right. \nonumber\\
&&\qquad+\left.\left\{\int d^2 \theta\left(\frac14f_a W^{a,\alpha}W^a_\alpha +C^3 W\right)+{\rm h.c.}\right\}\right],\label{eq:genesugra}
\end{eqnarray}
where $C=C_0+\theta\theta F^C$ is the chiral compensator superfield 
and the metric $g^C$ is defined by 
$g_{\mu\nu}^C = (C\bar C)^{-1}e^{K/3}g_{\mu\nu}^E$ for the Einstein-frame metric $g_{\mu\nu}^E$. 
Our obtained action is given in a so-called string frame 
and we can choose $C_0 = e^{-\phi_4}e^{K/6}$ to arrive at the frame in 
the above conformal SUGRA, 
where the VEV of the 4D dilaton $\phi_4$ is determined as 
\begin{equation*}
e^{-2\langle\phi_4\rangle} = e^{-2\langle\phi_{10}\rangle} \prod_i\mathcal A^{(i)} 
= g^{-4}\prod_i\mathcal A^{(i)}. 
\end{equation*} 
The K\"ahler potential for the moduli fields is given by 
\begin{equation*}
K^{(0)}  = - {\rm log} \left( S+\bar S\right) 
- {\rm log} \prod_{i=1}^3\left( T^{(i)}+\bar T^{(i)}\right)
- {\rm log} \prod_{i=1}^3\left( U^{(i)}+\bar U^{(i)}\right). 
\end{equation*}

When we compare the obtained action (\ref{eq:obtain}) 
with the general SUGRA action (\ref{eq:genesugra}) in the string frame, 
the K\"ahler potential $K$, the superpotential $W$ and the gauge kinetic function $f_a$ 
in the conformal SUGRA formulation can be identified as 
\begin{eqnarray}
K &=& K^{(0)} +\sum_{i,j} \sum_{a,b} \sum_{\mathcal I_{ab}} Z^{\bar ij}_{\mathcal I_{ab}}
{\rm Tr}\left[\bar\phi_{\bar i}^{\mathcal I_{ab}} e^{-V^a} \phi_j^{\mathcal I_{ab}} e^{V^b}\right],
\nonumber\\
W &=& \sum_{i,j,k} \sum_{a,b,c} \sum_{\mathcal I_{ab},\mathcal I_{bc},\mathcal I_{ca}} 
\lambda^{ijk}_{\mathcal I_{ab}\mathcal I_{bc}\mathcal I_{ca}} {\rm Tr}
\left[\phi_i^{\mathcal I_{ab}}\phi_j^{\mathcal I_{bc}}\phi_k^{\mathcal I_{ca}}\right], 
\nonumber\\
f_a &=& S, \label{eq:kineticg}
\end{eqnarray}
 where the K\"ahler metric $Z_{\mathcal I_{ab}^{\bar ij}}$ and the holomorphic Yukawa 
coupling $\lambda^{ijk}_{\mathcal I_{ab}\mathcal I_{bc}\mathcal I_{ca}}$ are given by 
\begin{eqnarray*}
Z_{\mathcal I_{ab}^{\bar ij}} &=& e^{2\langle\phi_4\rangle}\tilde Z_{\mathcal I_{ab}^{\bar ij}}, \\
\lambda^{ijk}_{\mathcal I_{ab}\mathcal I_{bc}\mathcal I_{ca}} &=& 
e^{3\langle\phi_4\rangle}e^{-K^{(0)}/2} 
\tilde\lambda^{ijk}_{\mathcal I_{ab}\mathcal I_{bc}\mathcal I_{ca}}. 
\end{eqnarray*}
$\tilde Z_{\mathcal I_{ab}^{\bar ij}}$ and 
$\tilde\lambda^{ijk}_{\mathcal I_{ab}\mathcal I_{bc}\mathcal I_{ca}}$ 
have been defined in Eqs.~(\ref{eq:tildez}) and (\ref{eq:tildelambda}), respectively. 

These should be shown as functions of the only moduli fields and 
an additional manipulation is required for that. 
If we promote straightforwardly the parameters to the moduli fields 
in accordance with Eq.~(\ref{eq:moduli}) in the above expressions, 
the Yukawa couplings will contain both the chiral and anti-chiral superfields 
and the holomorphicity of the superpotential is broken. 
Correct combinations of these parameters should be promoted to the moduli fields 
in the superpotential and the rest must be removed from the superpotential to 
the K\"ahler potential by rescaling the superfields $\phi_i^{\mathcal I_{ab}}$.

We consider the following rescaling \footnote{ 
This paper shows the explicit rescaling rules for the chiral fields, which 
determines the moduli dependence of their K\"ahler metrics. 
We should note that it is not completely deterministic. 
Indeed, there are many ways of the rescaling to remove the ill-defined factors, 
and we show the most plausible one. 
This discussion was also done in Ref.~\cite{DiVecchia:2008tm}. }, 
\begin{equation*}
\phi_i^{\mathcal I_{ab}} \rightarrow\alpha_{ab}^{(i)}\phi_i^{\mathcal I_{ab}}, 
\end{equation*}
where 
\begin{eqnarray*}
\alpha_{ab}^{(i)} &=& \frac1{g\sqrt{2\im\tau_i}}\left(\prod_r
\frac{\mathcal A^{(r)}}{\sqrt{2\im\tau_r}}\right)^{1/2}\\
&&\qquad \times {\rm exp}
\left[-\sum_r\frac{\pi i}{\im\tau_r}\frac{\bar\zeta_{ab}^{(r)}}{M_{ab}^{(r)}}\im\zeta_{ab}^{(r)}\right] 
\left(\frac{|M_{ab}^{(i)}|}{\prod_{r\neq i}|M_{ab}^{(r)}|}\right)^{1/4}, 
\end{eqnarray*}
and we promote the remaining parameters to the moduli fields after this. 
As the result, we can obtain the moduli depending form of 
the K\"ahler metric $Z_{\mathcal I_{ab}^{\bar ij}}$ 
and the holomorphic Yukawa coupling 
$\lambda^{ijk}_{\mathcal I_{ab}\mathcal I_{bc}\mathcal I_{ca}}$ as follows, 
\begin{eqnarray*}
Z_{\mathcal I_{ab}^{\bar ij}} &=& \delta^{\bar ij}
\left(\frac{T_j+\bar T_{\bar j}}{2}\right)^{-1}
\left(\prod_{r=1}^3\frac{U_r+\bar U_{\bar r}}{2}\right)^{-1/2}\\
&&\qquad
\frac1{2^{5/2}}
\left(\frac{|M_{ab}^{(j)}|}{\prod_{r\neq j}|M_{ab}^{(r)}|}
\right)^{1/2}
{\rm exp}\left[-\sum_{r=1}^3\frac{4\pi}{U_r+\bar U_{\bar r}}
\frac{\left(\im\zeta_{ab}^{(r)}\right)^2}{M_{ab}^{(r)}}\right], \\
\lambda^{ijk}_{\mathcal I_{ab}\mathcal I_{bc}\mathcal I_{ca}} &=& 
-\frac13\epsilon^{\rm ijk}\delta_{\rm i}^i\delta_{\rm j}^j\delta_{\rm k}^k
\prod_{r=1}^3\lambda^{(r)}_{I_{ab}^{(r)} I_{bc}^{(r)} I_{ca}^{(r)}}. 
\end{eqnarray*}
where 
\begin{equation}
\lambda^{(r)}_{ I_{ab}^{(r)} I_{bc}^{(r)} I_{ca}^{(r)}} = 
\left\{ \begin{array}{ll}
\displaystyle 
\lambda^{(r)}_{ab,c} & \quad (M^{(i)}_{ab} > 0) \\
\displaystyle 
\lambda^{(r)}_{bc,a} & \quad (M^{(i)}_{bc} > 0) \\
\displaystyle 
\lambda^{(r)}_{ca,b} & \quad (M^{(i)}_{ca} > 0) 
\end{array} \right. \label{eq:yukafinxx}
\end{equation}
and 
\begin{eqnarray}
\lambda^{(r)}_{ab,c} &=& 
\sum_{m=1}^{M^{(r)}_{ab}} 
\delta_{I^{(r)}_{bc}+I^{(r)}_{ca}-m M^{(r)}_{bc},~I^{(r)}_{ab}}\nonumber\\
 && \times 
\vartheta 
\begin{bmatrix}
\frac{
M^{(r)}_{bc} I^{(r)}_{ca} - M^{(r)}_{ca} I^{(r)}_{bc} + m M^{(r)}_{bc} M^{(r)}_{ca}}
{M^{(r)}_{ab} M^{(r)}_{bc} M^{(r)}_{ca}} 
\\ 0 \end{bmatrix} 
\left( 
 \bar\zeta^{(r)}_{ca}M^{(r)}_{bc} - \bar\zeta^{(r)}_{bc}M^{(r)}_{ca}, 
iU_r M^{(r)}_{ab}M^{(r)}_{bc}M^{(r)}_{ca} \right)~. \label{eq:finalyukaholo}
\end{eqnarray}
These expressions are valid for $M^{(r)}_{ab}M^{(r)}_{bc}M^{(r)}_{ca}>0$. 
The study with vanishing magnetic fluxes $M^{(r)}_{ab}M^{(r)}_{bc}M^{(r)}_{ca}=0$ 
is also shown in Ref.~\cite{Abe:2012ya}. 

\section{6D and 10D SYM theories and their mixtures} 
\label{sec:6d10d}
In this section, we give extensions of the previous work given in the 10D SYM theory. 
We can expect the way of dimensional reduction and obtaining the 4D effective SUGRA action 
to be applied to the $(4+2n)$-dimensional cases ($n=1,2,3$) and their mixtures, 
because such SYM systems can be derived 
from single 10D SYM theories through ``partial" dimensional reductions. 

\subsection{Superfield description of the 6D and 10D SYM theories}
For a while, we concentrate on an instructive case which consists of 
a six-dimensional (6D) SYM theory and a 10D SYM theory. 

6D SYM theories with a vector multiplet and a hyper multiplet of 
4D $\mathcal N=2$ SUSY are straightforwardly obtained  by 
dimensional reductions of 10D SYM theories. 
In the 10D SYM theories compactified on a flat space without magnetic fluxes and so on,  
the zero-mode wavefunctions of the 10D fields are given by a constant in the space, 
and we can then perform integrations of the action with respect to the flat directions. 
For example, when we perform the integration with respect to 
four extra-dimensional coordinates $(x_6, x_7, x_8, x_9)$, 
that induces just the 4D volume factor and a 6D effective action is directly derived. 
The 6D vector $A_m$ $m=0,1,\ldots,5$ and a part of the 10D 
Majorana-Weyl spinor form an $\mathcal N=2$ vector multiplet, 
and the other parts form an $\mathcal N=2$ hyper multiplet. 

In mixtures of 6D SYM theory and 10D SYM theory, 
there should appear an additional mixing sector. 
This consists of bifundamental representations 
which are charged under both the 6D and 10D SYM theories. 
They form another hyper multiplet because the mixing part also has 
the $\mathcal N=2$ SUSY counted by the 4D supercharges. 
Furthermore, since they are coupled to the 10D gauge fields as well as the 6D gauge fields, 
it is sensible for their wavefunctions to depend on all of  the 10D coordinates 
but have profiles of point-like quasi-localizations 
in the four $(x_6, x_7, x_8, x_9)$-directions. 

The positions of localization points are significant because 
they are related to the magnitude of their coupling constants in the 4D effective theory. 
The positions are determined by the VEVs of position moduli, 
and field contained in the hyper multiplet plays the role in our SYM systems. 

These situations are realized or understood in single 10D SYM theories 
by introducing infinite magnetic fluxes in the four directions~\cite{Cremades:2004wa}. 
To demonstrate, let us consider a 10D $U(M+N)$ SYM theory compactified on three tori, 
and introduce an infinite magnetic flux on two of the three tori to 
break the gauge group as $U(M+N)\rightarrow U(M)\times U(N)$. 
In this scheme, adjoint representations of the unbroken subgroups $U(M)$ and $U(N)$ 
do not feel the magnetic fluxes, thus they have a flat zero-mode profile. 
Carrying out the integration of the $U(M)$ SYM action 
(either of the two SYM actions) on the two infinitely magnetized tori 
leads to the 6D $U(M)$ SYM theory and the other is still the 10D $U(N)$ SYM theory. 
Bifundamental representations $(M,\bar N)$ and $(\bar M, N)$ feel then 
the infinite magnetic fluxes which localize them at a point on the two tori. 

To see the details, we consider the limit $|M|\rightarrow \infty$ in the following 
integral of zero-mode wavefunctions of the bifundamentals, 
\begin{equation*}
\int_{T^2} d(\re z) \left(\Theta^{I,M}\right)^*\Theta^{I,M} 
= 
\frac{(2\im\tau |M|)^{1/2}}{\mathcal A}
\sum_n e^{-2\pi |M|\im\tau
\left(n+\frac I{|M|}+\frac{\im z}{\im\tau}\right)^2},  
\end{equation*}
which appears in the way of the normalization (\ref{eq:normal}) 
and the wavefunction $\Theta^{I,M}$ is defined in Eq.~(\ref{eq:waveform}). 
In the limit of infinite magnetic fluxes, this integral gives a delta function as 
\begin{equation}
\int_{T^2} d(\re z) \left(\Theta^{I,M}\right)^*\Theta^{I,M} 
= 
\frac1{\mathcal A}
\sum_n \delta\left(\frac{\im z}{\im\tau}+n+\frac I{|M|}\right) 
\label{eq:deltadefss}. 
\end{equation}
The infinite magnetic fluxes induce an infinite number of the zero-modes 
labeled by ``$I$". 
They are quasi-localized at different points with an interval $1/M$. 
That is, this torus is filled up with an infinite number of zero-modes 
but each of which is localized at different points like the delta function. 
Now, we choose a zero-mode with $I=0$, 
which zero-mode is quasi-localized at the origin on the torus, 
and eliminate the other zero-modes by hand\footnote{ 
This will not break the SUSY.}. As a result, the desirable bifundamental representation 
is obtained. The summation of ``$n$" gives the delta function to a certain periodicity on the torus 
and the right-hand side of Eq.~(\ref{eq:deltadefss}) with $I=0$ can be 
identified as a well-defined delta function on the torus.  
We denote it by $\delta_{T^2}(z)$ as is used in Ref.~\cite{Cremades:2004wa}, 
that is, 
\begin{equation}
\delta_{T^2}(z)\equiv\frac1{\mathcal A}
\sum_n \delta\left(\frac{\im z}{\im\tau}+n\right).\label{eq:tdelta}
\end{equation} 

We can infer from this result that the point-like localization 
of the bifundamental representations is described by $\Theta^{0,M}$ with 
\begin{equation*}
\Theta^{0,M}\sim \sqrt{\delta_{T^2}(z)}, 
\end{equation*}
which is caused by the infinite magnetic flux $M$. 
When we consider Wilson lines on the magnetized torus, this is translated as 
\begin{equation*}
\sqrt{\delta_{T^2}(z)}\rightarrow\sqrt{\delta_{T^2}(z+\zeta)}. 
\end{equation*} 
Since the Wilson lines on the torus is given as the VEVs of the fields contained 
in the hyper multiplet, those fields can be identified with the position moduli fields 
as we expected. 

In the rest of this subsection, we derive a specific form of the effective action corresponding to 
the mixture of  the 6D $U(M)$ SYM theory compactified on $(T^2)_1$ 
and the 10D $U(M)$ SYM theory on $(T^2)_1\times (T^2)_2\times (T^2)_3$, 
by introducing the following infinite magnetic fluxes in a 10D $U(M+N)$ SYM theory,  
in accordance with the vacuum configuration (\ref{eq:aconfig}), 
\begin{eqnarray} 
M^{(1)}&=&\begin{pmatrix}
0\times{\bm 1}_{M}&0\\
0&0\times{\bm 1}_{N}
\end{pmatrix},\nonumber\\
M^{(2)}=\begin{pmatrix}
H\times{\bm 1}_{M}&0\\
0&0\times{\bm 1}_{N}
\end{pmatrix}, & &\qquad
M^{(3)}=\begin{pmatrix}
-H\times{\bm 1}_{M}&0\\
0&0\times{\bm 1}_{N}
\end{pmatrix},\label{eq:fluxinft}
\end{eqnarray}
where we take the limit $H\rightarrow\infty$. 
These matrices represent the internal space of $U(M+N)$. 
In the VEV of the form (\ref{eq:aconfig}), 
we can also introduce the Wilson lines $\zeta^{(i)}$ and 
they shift the point-like localized wavefunctions of 
bifundamental representations $(M,\bar N)$ and $(\bar M, N)$ 
by $\zeta^{(i)}_{MN}/H$ ($\zeta^{(i)}_{MN}\equiv \zeta^{(i)}_{M}-\zeta^{(i)}_{N}$). 
That is, the wavefunctions are shifted as 
\begin{equation*}
\sqrt{\delta_{T^2}(z)}\rightarrow\sqrt{\delta_{T^2}(z+\zeta^{(i)}_{MN}/H)}. 
\end{equation*} 
This deviation vanishes in the limit $H\rightarrow\infty$ unless the Wilson lines 
$\zeta^{(i)}_{MN}$ given by the position moduli  take infinite values. 
This is one of differences between the usual Wilson lines and the VEVs of the position moduli.

These infinite magnetic fluxes $H$ and $-H$ 
induce a kind of chirality projection as well as the point-like localizations. 
As the result, some of the zero-modes are eliminated as 
\begin{eqnarray*}
V&=&\begin{pmatrix}
V^m&0\\
0&V^n
\end{pmatrix},\\
\phi_1=\begin{pmatrix}
\phi_1^m&0\\
0&\phi_1^n
\end{pmatrix},\qquad
\phi_2&=&\begin{pmatrix}
\phi_2^m&g\phi_2^{mn}\\
0&\phi_2^n
\end{pmatrix},\qquad
\phi_3=\begin{pmatrix}
\phi_3^m&0\\
g\phi_3^{mn}&\phi_3^n
\end{pmatrix}. 
\end{eqnarray*}
Now, we assign that the first-block entries $V^m$ and $\phi_i^m$ to 
the 6D $U(M)$ SYM theory and the last-block entries $V^n$ and $\phi_i^n$ 
to the 10D $U(N)$ SYM theory. 
The $U(N)$ part labeled by $``n"$ is the same as is reviewed in the previous section. 
The 6D gauge fields $V^m$ and $\phi_1^m$ form an $\mathcal N=2$ vector multiplet. 
A hyper multiplet is also composed of the fields $\phi_2^m$ and $\phi_3^m$, 
which are identified with the position moduli. 
The bifundamentals $\phi_2^{mn}$ and $\phi_3^{mn}$ form another hyper multiplet 
and their action will be given to have an $SU(2)_{\rm R}$ invariance. 
Here they are normalized by the gauge coupling constant $g$ for later convenience.

In the superfield description of the 10D $U(M+N)$ SYM theory, 
we consider the infinitely magnetized background to obtain 
6D and 10D pure SYM theories. 
The action is composed of three parts as follows, 
\begin{equation}
\mathcal S = S_{m}+S_{n}+S_{mn}. \label{eq:sss}
\end{equation}
First, the explicit form of the 6D SYM action $S_{m}$ is obtained 
by the dimensional reduction for the second and the third tori. 
Since the relevant fields $V^m$ and $\phi_i^m$ do not feel the magnetic fluxes 
and their wavefunctions are flat on the two tori, 
the dimensional reduction can be straightforwardly performed. 
As the result, we find 
\begin{equation}
S_m=\frac{\mathcal A^{(2)}\mathcal A^{(3)}}{g^2}
\int d^6X\sqrt{-G_6}\int d^4\theta \mathcal K_m +\left\{
\int d^2\theta\left(\frac1{4}\mathcal W_m^\alpha
{\mathcal W_\alpha}_m
+\mathcal W_m \right)+{\rm h.c.}\right\},\nonumber
\end{equation}
where $G_6$ is the determinant of the 6D spacetime metric, $M^4\times (T^2)_1$, 
and the three functions $\mathcal K_m$, $\mathcal W_m$ and 
$\mathcal W_m^\alpha$  are given by 
\begin{eqnarray}
\mathcal K_m &=& 2{\rm Tr}\left[
h^{11}\left(\left(\sqrt2\bar\partial_1+\bar\phi_1^m\right)e^{-V^m}\right)
\left(-\sqrt2\partial_1+\phi_1^m\right)e^{V^m}\right.\nonumber\\
& & \hspace{20pt}+h^{11}\bar\partial_1e^{-V^m}\partial_1e^{V^m}
+h^{22}\bar\phi_2^m e^{-V^m}\phi_2^m e^{V^m}\nonumber\\
& & \hspace{20pt}\left.+h^{33}\bar\phi_3^m e^{-V^m}\phi_3^m e^{V^m}
+\mathcal K'_{\rm WZW}\right],\nonumber\\
\mathcal W_m &=& 2\sqrt2 \left(e_1e_2e_3\right)^{-1}\phi_3^m
\left(\partial_1\phi_2^m-\frac1{\sqrt2}\left[\phi_1^m, \phi_2^m\right]\right),\nonumber\\
\mathcal {W_\alpha}_m &=& -\frac14\bar D\bar De^{-V^m}D_\alpha e^{V^m}. 
\nonumber
\end{eqnarray}
The determinant of the vielbein $e_i$ is given by $\sqrt2\left(2\pi R_i\right)$ and 
the derivative terms with respect to $z_2$ and $z_3$ vanishes 
because of the flat wavefunctions. 
In this dimensional reduction, 
we adopt a normalization where the flat zero-mode wavefunctions are given by 1, 
instead of Eq.~(\ref{eq:normal}). Thus, the integration 
just induces a global factor corresponding to the volume $\mathcal A^{(2)}\mathcal A^{(3)}$. 
Although the prefactor $\mathcal A^{(2)}\mathcal A^{(3)}/g^2$ seems 
to be a gauge coupling constant of this 6D $U(M)$ SYM theory, 
we should replace this by a new symbol as 
$\mathcal A^{(2)}\mathcal A^{(3)}/g^2\rightarrow 1/g_6^2$ 
because this can be generically independent of the volume 
of the other four extra dimensions of space in pure 6D theories.

Next, we consider the 10D $U(N)$ SYM part $S_n$ in Eq.~(\ref{eq:sss}). 
This part is elicited from the original 10D $U(M+N)$ SYM theory directly: 
\begin{equation*}
S_n=\int d^{10}X\sqrt{-G}\int d^4\theta \mathcal K_n +\left\{
\int d^2\theta\left(\frac1{4g^2}\mathcal W_n^\alpha
{\mathcal W_\alpha}_n
+\mathcal W_n\right)+{\rm h.c.}\right\},
\end{equation*}
where the three function $\mathcal K_n$, $\mathcal W_n$ and 
$\mathcal W_n^\alpha$ of superfields are given by 
\begin{eqnarray*}
\mathcal K_n &=& \frac2{g^2}h^{i\bar j}{\rm Tr}\left[
\left(\sqrt2\bar\partial_{\bar i}+\bar\phi^n_{\bar i}\right)e^{-V^n}
\left(-\sqrt2\partial_j+\phi^n_j\right)e^{V^n}
+\bar\partial_{\bar i}e^{-V^n}\partial_je^{V^n}\right]+\mathcal K_{\rm WZW},\nonumber\\
\mathcal W_n &=& \frac1{g^2}\epsilon^{\rm ijk}e_{\rm i}^{~i}e_{\rm j}^{~j}
e_{\rm k}^{~k} {\rm Tr}\left[\sqrt2\phi^n_i\left(\partial_j\phi^n_k-\frac1{3\sqrt2}
\left[\phi^n_j, \phi^n_k\right]\right)\right],\nonumber\\
\mathcal {W_\alpha}_n &=& -\frac14\bar D\bar De^{-V^n}D_\alpha e^{V^n}. 
\end{eqnarray*}
This has the same form as that of the original 10D SYM action. 

In the last part $S_{mn}$, 
the infinite magnetic flux is a key to derive effective actions, 
which is analogous to an off-diagonal part of 10D SYM theories 
shown in the previous section. 
Substituting the vacuum configuration ($\ref{eq:fluxinft}$) for 
$\langle\phi_i\rangle$ and $\langle\bar\phi_{\bar i}\rangle$ 
in the action (\ref{eq:symmag}), 
the zero-mode equations for $\phi_2^{mn}$ and $\phi_3^{mn}$ on the second and the third tori 
are given by 
\begin{eqnarray}
\left[\bar\partial_{\bar i} +\frac{\pi}{2\im\tau_i}
\left(H z_i+\zeta_{MN}^{(i)}\right)\right](f_j^{(i)})^{mn} &=& 0 \qquad{\rm for}\quad i=j,
\label{eq:zeroequhii}\\
\left[\partial_i -\frac{\pi}{2\im\tau_i}
\left(H\bar z_{\bar i}+\bar\zeta_{MN}^{(i)}\right)\right]
(f_j^{(i)})^{mn} &=& 0 \qquad{\rm for}\quad i\neq j, \label{eq:zeroequhij}
\end{eqnarray}
where $i,j=2,3$ and $(f_j^{(i)})^{mn}$ represents the zero-mode wavefunction of 
$\phi_j^{mn}$ on the $i$-th torus omitting the mode number $n_i=0$. 
We also consider the Wilson lines in addition to the infinite magnetic fluxes here 
as we discussed it in the below of Eq.~(\ref{eq:fluxinft}). 
These equations have an infinite numbers of normalizable solutions 
labeled by an index $I^{(i)}_{mn}$ in the limit $H\rightarrow\infty$, 
and we pick up one of them by $I^{(i)}_{mn}=0$. 
Thus, we obtain the well-defined delta function (\ref{eq:tdelta}) which expresses 
the point-like localization as the solution of the zero-mode equation. 
We can carry out the integration with respect to $z_2$ and $z_3$ 
in the action and obtain the following form, 
\begin{eqnarray}
S_{mn} &=& \int~d^{6}X\sqrt{-G_6}\int~d^4\theta {~\rm Tr~}
\left(2h^{22}\bar\phi_2^{mn}e^{-V^m}\phi_2^{mn} e^{V^n}
+2h^{33}\phi_3^{mn}e^{V^m}\bar\phi_3^{mn}e^{-V^n}\right)\nonumber\\
&&\hspace{-5pt}+2\sqrt2\left(e_1e_2e_3\right)^{-1} \int~d^2\theta {~\rm Tr~}
\left[\phi_3^{mn}\left(\partial_1\phi_2^{mn}-\frac 1{\sqrt2}\phi_1^m\phi_2^{mn}
+\frac Q{\sqrt2}\phi_2^{mn}\phi_1^n\right) +{\rm h.c.}\right], \label{eq:smn6d} 
\end{eqnarray}
where the factor $Q$ is given by the integrals on the two tori as, 
\begin{eqnarray}
Q&=&\prod_{s=2,3}\int dz^sd\bar z^{\bar s} 
\left\{(f_1^{(s)})^{n}(z_s)\times \delta_{T^2}(z_s+\tilde\zeta^{(s)})\right\}, \nonumber\\
&=&\prod_{s=2,3} (f_1^{(s)})^{n} (\tilde\zeta^{(s)}), \label{eq:afactor}
\end{eqnarray}
with $\tilde\zeta^{(s)}\equiv\zeta_{MN}^{(s)}/H$. 
In the above, $(f_1^{(s)})^{n}(\tilde\zeta^{(s)})$ is the zero-mode wavefunction of $\phi_1^n$ 
on the $s$-th torus and its constant argument $\tilde\zeta^{(s)}$ represents the position of 
the point-like quasi-localization on the tori. 
(Note that this zero-mode wavefunction $(f_1^{(s)})^n$ is 
a constant function now because physical finite fluxes are absent in Eq.~(\ref{eq:fluxinft}).) 

We have obtained the superfield description of the mixture of the 6D $U(M)$ SYM theory and 
the 10D $U(N)$ SYM theory, which is derived from the 10D $U(M+N)$ SYM theory by 
introducing the infinite magnetic fluxes. 

\subsection{4D effective action on magnetized backgrounds } 
The infinite magnetic fluxes have yielded the action for the 6D $U(M)$ SYM theory, 
the 10D $U(N)$ SYM theory and their mixing part compactified on 
``virtually pure" tori. 
That is, the infinite magnetic fluxes are used for only realizing the point-like localizations and 
inducing a kind of projections, 
and they lead to un-magnetized higher-dimensional SYM systems. 
In the following, we add ``finite (physical)" magnetic fluxes in this mixture of 
the SYM theories. 
We consider the following configuration of magnetic fluxes, 
instead of Eq.~(\ref{eq:fluxinft}), 
\begin{eqnarray}
M^{(1)}&=&\begin{pmatrix}
M^{(1)}_m &0\\
0&M^{(1)}_n
\end{pmatrix},\nonumber\\
M^{(2)}=\begin{pmatrix}
M_m^{(2)}+H\times{\bm 1}_{M}&0\\
0&M^{(2)}_n
\end{pmatrix}, & &\qquad
M^{(3)}=\begin{pmatrix}
M_m^{(3)}-H\times{\bm 1}_{M}&0\\
0&M^{(3)}_n
\end{pmatrix},\label{eq:flux610}
\end{eqnarray} 
where the $(M\times M)$-matrix $M^{(i)}_m$ and 
the $(N\times N)$-matrices $M^{(i)}_n$ represent finite magnetic fluxes, 
and the infinite magnetic flux is also introduced by $H$ in the limit $H\rightarrow \infty$. 

This is a generic form of flux configurations. Some of their entries would have some constraints. 
For instance, 
the matrices $M^{(2)}_m$ and $M^{(3)}_m$ should be restricted not to break 
the gauge symmetry, that is,  $M^{(2)}_m\propto M^{(3)}_m\propto {\bm 1}_{M}$, 
otherwise the zero-mode wavefunctions of the 6D fields are deformed 
and their spectrum is shifted by structure of the 4D extra space 
which is not related to the 6D SYM theory. 
When $M^{(2)}_m\propto M^{(3)}_m\propto {\bm 1}_{M}$, 
the matrix $M^{(1)}_m$ should also be proportional to the identity to preserve 
the $\mathcal N=1$ SUSY. 
Note that these trivial magnetic fluxes in the $U(M)$ sector $M^{(i)}_m\propto {\bm 1}_{M}$ 
can be eliminate by a shift of flux configurations as 
$M^{(i)}\rightarrow M^{(i)}+m_i\times{\bm 1}_{M+N}$ 
because the two configurations in the SYM theories lead to 
equivalent 4D effective theories. 
For completeness of our description, 
we consider a general form of $M^{(1)}_m$ in the following calculations 
even if it breaks the $\mathcal N=1$ SUSY. 
The others $M^{(2)}_m$ and $M^{(3)}_m$ are proportional to ${\bm 1}_{M}$, 
and they have no affect on the 4D effective theory in the limit $H\rightarrow \infty$. 
The 4D effective action with $M^{(1)}_m\propto{\bm 1}_{M}$ is also calculated in Appendix A. 

When some of diagonal entries of the matrices $M^{(1)}_m$ and $M^{(i)}_n$ 
take degenerate values, 
the two gauge symmetries are broken as 
$U(M)\rightarrow\prod_{a'} U(M_{a'})$ and $U(N)\rightarrow\prod_{a} U(N_{a})$. 
The unbroken gauge subgroups of the 6D $U(M)$ SYM theory are labeled by 
indices $a',b',c'$. The indices $a,b,c$ label the remaining subgroups 
of the 10D $U(N)$ SYM theory and this is the same as in the previous section. 

We discuss the zero-mode equations and wavefunctions 
on this magnetized background. 
Since those obtained in the 10D $U(N)$ SYM sector and 
its 4D effective SUGRA action are given in the previous section with the same notation, 
we focus on the other sectors. 
First, we consider the 6D $U(M)$ SYM part $S_m$ which contains only fields 
with the subscript $m$, 
the superfield description of which on the magnetized background is given by 
\begin{equation}
S_m=\frac1{g_6^2}
\int d^6X\sqrt{-G_6}\int d^4\theta \mathcal K_m +\left\{
\int d^2\theta\left(\frac1{4}\mathcal W_m^\alpha
{\mathcal W_\alpha}_m
+\mathcal W_m \right)+{\rm h.c.}\right\},\nonumber
\end{equation}
where the three functions $\mathcal K_m$, $\mathcal W_m$ and 
$\mathcal W_m^\alpha$  are given by 
\begin{eqnarray}
\mathcal K_m &=& 2 h^{\bar ij}{\rm Tr} \left[
\bar\phi_{\bar i}^m\phi_j^m+[\bar\phi_{\bar i}^m,\,\phi_j^m]V^m 
+\left(\bar\partial_{\bar i}V^m\right)\left(\partial_jV^m\right) 
+\frac12\left(\bar\phi_{\bar i}^m\phi_j^m+\phi_j^m\bar\phi_{\bar i}^m\right)(V^m)^2
-\bar\phi_{\bar i}^mV^m\phi_j^mV^m\right] \nonumber\\
&& 
+2 \sqrt2 h^{\bar 11}{\rm Tr} \left[\left(\bar\partial_{\bar 1}\phi_1^m+\frac1{\sqrt2}
[\langle\bar\phi_{\bar 1}^m\rangle,\,\phi_1^m] + {\rm h.c.} \right)V^m\right] 
+\mathcal K_m^{(\rm D)},\nonumber\\
\mathcal W_m &=& 2\sqrt2 \left(e_1e_2e_3\right)^{-1}\phi_3^m
\left(\partial_1\phi_2^m-\frac1{\sqrt2}\left[\langle\phi_1^m\rangle, \phi_2^m\right]
-\frac1{\sqrt2}\left[\phi_1^m, \phi_2^m\right]\right) 
+\mathcal W_m^{(\rm F)}.\nonumber
\end{eqnarray}

In the assumption of $U(M)$ gauge symmetry breaking due to the magnetic fluxes $M_m^{(1)}$,  
we derive the zero-mode equations for the relevant fields $(\phi_i^{m})_{a'b'}$ 
on the torus $(T^2)_1$ from this action, which are described 
as follows, 
\begin{eqnarray*}
\left[\bar\partial_{\bar 1} +\frac{\pi}{2\im\tau_1}
\left(M_{a'b'}^{(1)}z_1+\zeta_{a'b'}^{(1)}\right)\right](f_1^{(1)})^m_{a'b'} &=& 0, \\
\left[\partial_1 -\frac{\pi}{2\im\tau_1}
\left(M_{a'b'}^{(1)}\bar z_{\bar 1}+\bar\zeta_{a'b'}^{(1)}\right)\right]
(f_i^{(1)})^m_{a'b'} &=& 0\qquad {\rm for~}i=2,3,\\
\end{eqnarray*}
where $(f_i^{(1)})^m_{a'b'}$ represents the zero-mode wavefunction of 
bifundamental $(\phi_i^{m})_{a'b'}$ on the first torus, 
and the magnetic fluxes and the Wilson lines are defined as 
$M_{a'b'}^{(1)}\equiv (M_m^{(1)})_{M_{a'}} - (M_m^{(1)})_{M_{b'}}$ and 
$\zeta_{a'b'}^{(1)}\equiv (\zeta_m^{(1)})_{M_{a'}} - (\zeta_m^{(1)})_{M_{b'}}$. 
This is similar to those in Eqs.~(\ref{eq:zeroii}) and (\ref{eq:zeroij}). 
When the sign of the magnetic fluxes is correctly chosen, 
we can obtain $|M_{a'b'}^{(1)}|$ normalizable solutions labeled by the index 
$I_{a'b'}=1,2,\ldots, |M_{a'b'}^{(1)}|$. 

We describe the zero-modes in the 4D effective action as follows, 
\begin{equation*}
(V^{m,n_1=0})_{a'a'}\equiv V^{a'}, 
\qquad 
(\phi_i^{m,n_1=0})_{a'b'}\equiv g_6 \phi_i^{I_{a'b'}}.  
\end{equation*}
We use the similar notation to the previous section: 
$V^{a'}$ represents the zero-mode of an adjoint representation of $U(M_{a'})$ 
and $\phi_i^{a'b'}$ is the zero-mode of a bifundamental one $(M_{a'},~\bar M_{b'})$. 
We can omit the subscript $m$ because 
they have the YM indices $a'b'$ which represent the gauge subgroups of $U(M)$. 
The adjoint representation $V^{a'}$ 
do not feel the magnetic fluxes and their zero-modes have a trivial profile. 
We calculate the 4D effective action in the same manner and find 
\begin{eqnarray}
S_m=\int d^4 x \left[
\int d^4\theta {\mathcal K_{m,\rm eff}}+\left\{
\int d^2\theta\left(\frac14\mathcal W^{a',\alpha}_m\mathcal W^{a'}_{m,\alpha} 
+\mathcal W_{m,\rm eff}\right)+ {\rm h.c.}
\right\}\right] 
\end{eqnarray}
where the functions $\mathcal K_{m,\rm eff}$, $\mathcal W_{m,\rm eff}$ and 
$\mathcal W^{a'}_{m,\alpha}$ have the following form, 
\begin{eqnarray*}
\mathcal K_{m,\rm eff} &=& \sum_{i,j}\sum_{a',b'}\sum_{I_{a'b'}^{(1)}} 
\tilde Z_{I_{a'b'}}^{\bar i j} {\rm Tr} \left[ 
\bar\phi_{\bar i}^{I_{a'b'}} e^{-V^{a'}} \phi_j^{I_{a'b'}} e^{V^{a'}} \right], \\
\mathcal W_{\rm eff} &=& \sum_{i,j,k}\sum_{a',b',c'}
\sum_{I_{a'b'},I_{b'c'},I_{c'a'}}
\tilde\lambda^{ijk}_{I_{a'b'}I_{b'c'} I_{c'a'}}{\rm Tr} 
\left[\phi_i^{I_{a'b'}}\phi_j^{I_{b'c'}}\phi_k^{I_{c'a'}}\right],\\
\mathcal W^{a'}_\alpha &=& -\frac1{4g_{a'}^2}\bar D\bar D e^{-V^{a'}}D_\alpha e^{V^{a'}}.\qquad 
g_{a'} = g_6 {\mathcal A^{(1)}}^{-1/2}~.
\end{eqnarray*}
In these expressions, the K\"ahler metric $\tilde Z_{I_{a'b'}}^{\bar i j}$ 
and holomorphic Yukawa coupling 
$\tilde\lambda^{ijk}_{I_{a'b'} I_{b'c'} I_{c'a'}}$ 
are determined by integrals in the 6D extra compact space 
and they can be written as 
\begin{eqnarray}
\tilde Z_{I_{a'b'}}^{\bar i j} &=& 2h^{\bar ij}\label{eq:6dtildezz}\\
\tilde\lambda^{ijk}_{I_{a'b'}I_{b'c'}I_{c'a'}} &=& 
-\frac{2g_6}3\epsilon^{\rm ijk}e_{\rm i}^{~i}e_{\rm j}^{~j}e_{\rm k}^{~k}  
\tilde\lambda^{(1)}_{I_{a'b'} I_{b'c'} I_{c'a'}}\label{eq:6dtildell},
\end{eqnarray}
where
\begin{equation*}
\tilde\lambda^{(1)}_{I_{a'b'} I_{b'c'} I_{c'a'}} = 
\left\{ \begin{array}{ll}
\displaystyle 
\tilde\lambda^{(1)}_{a'b',c'} & \quad (M^{(1)}_{a'b'} > 0) \\
\displaystyle 
\tilde\lambda^{(1)}_{b'c',a'} & \quad (M^{(1)}_{b'c'} > 0) \\
\displaystyle 
\tilde\lambda^{(1)}_{c'a',b'} & \quad (M^{(1)}_{c'a'} > 0) 
\end{array} \right., 
\end{equation*}
and 
\begin{eqnarray*}
\tilde\lambda^{(1)}_{a'b',c'} &=& 
{\mathcal N}_{M^{(1)}_{a'b'}}^{-1}{\mathcal N}_{M^{(1)}_{b'c'}}{\mathcal N}_{M^{(1)}_{c'a'}}
\sum_{m=1}^{M^{(1)}_{a'b'}} 
\delta_{I_{b'c'}+I_{c'a'}-m M^{(1)}_{b'c'},~I_{a'b'}}\\
 && \times 
\exp \left[ \frac{\pi i}{\im\tau_1} 
\left( 
  \frac{\bar\zeta^{(1)}_{a'b'}}{M^{(1)}_{a'b'}}\im\zeta^{(1)}_{a'b'}
+\frac{\bar\zeta^{(1)}_{b'c'}}{M^{(1)}_{b'c'}}\im\zeta^{(1)}_{b'c'}
+\frac{\bar\zeta^{(1)}_{c'a'}}{M^{(1)}_{c'a'}}\im\zeta^{(1)}_{c'a'}
\right) \right] \\
 && \times \vartheta 
\begin{bmatrix}
\frac{
M^{(1)}_{b'c'} I_{c'a'} - M^{(1)}_{c'a'} I_{b'c'} + m M^{(1)}_{b'c'} M^{(1)}_{c'a'}}
{M^{(1)}_{a'b'} M^{(1)}_{b'c'} M^{(1)}_{c'a'}} 
\\ 0 \end{bmatrix} 
\left( 
 \bar\zeta^{(1)}_{c'a'}M^{(1)}_{b'c'} - \bar\zeta^{(1)}_{b'c'}M^{(1)}_{c'a'}, 
-\bar\tau_1 M^{(1)}_{a'b'}M^{(1)}_{b'c'}M^{(1)}_{c'a'} \right)~. 
\end{eqnarray*}
The normalization factors are defined in Eq.~(\ref{eq:normal}). 

Next, we consider the mixing part $S_{mn}$, which consists of bifundamental representations 
$(M_{a'}, \bar N_{b})$ and their conjugate representations (Note that 
$U(M_{a'})$ and $U(N_{b})$ are subgroups of the gauge groups $U(M)$ and $U(N)$, 
respectively.). 
The infinite and finite magnetic fluxes (\ref{eq:flux610}), 
instead of Eq.~($\ref{eq:fluxinft}$), are introduced in the action (\ref{eq:symmag}). 
On the second and the third tori, the zero-mode equations 
$(\ref{eq:zeroequhii})$ and $(\ref{eq:zeroequhij})$ are a little modified by the finite fluxes, 
but the finite shift of $H$ does not affect in the limit $H\rightarrow \infty$. 
It leads to the same results on these two tori and 
the 6D action of the form (\ref{eq:smn6d}) is obtained again. 
In addition to that, we have the following zero-mode equations for 
$(\phi_2^{mn})_{a'b}$ and $(\phi_3^{mn})_{ab'}$ on the first torus, 
\begin{eqnarray*}
\left[\partial_{ 1} -\frac{\pi}{2\im\tau_1}
\left(M_{a'b}^{(1)}\bar z_1+\bar\zeta_{a'b}^{(1)}\right)\right](f_2^{(1)})^{mn}_{a'b} &=& 0,\\
\left[\partial_1 -\frac{\pi}{2\im\tau_1}
\left(M_{ab'}^{(1)}\bar z_{\bar 1}+\bar\zeta_{ab'}^{(1)}\right)\right]
(f_3^{(1)})^{mn}_{ab'} &=& 0,  
\end{eqnarray*}
where $(f_2^{(1)})^{mn}_{a'b}$ and $(f_3^{(1)})^{mn}_{ab'}$ are the zero-mode wavefunctions of 
the bifundamental representations $(\phi_2^{mn})_{a'b}$ and $(\phi_3^{mn})_{ab'}$, 
respectively. 
Note that $\phi_2^{mn}$ is the bifundamental representation $(M, \bar N)$ of 
the product gauge group $U(M)\times U(N)$. 
It contains only bifundamental representations as $(M_{a'}, \bar N_{b})$ 
and does not include the others $(\bar M_{a'}, N_{b})$. 
On the other hand, 
$\phi_3^{mn}$ is the bifundamental representation $(\bar M, N)$ which 
consists of only the bifundamental representations as $(\bar  M_{a'}, N_b)$. 

These zero-mode equations with the negative magnetic fluxes 
allow $|M_{a'b}^{(1)}|$ or $|M_{ab'}^{(1)}|$ normalizable solutions 
labeled by $I_{a'b}$ or $I_{ab'}$ in the same way as in Eq.~(\ref{eq:zeroij}).  
We express the corresponding zero-modes as 
\begin{equation*}
(\phi_2^{mn,n_1=0})_{a'b} \equiv \phi_2^{I_{a'b}},\qquad
(\phi_3^{mn,n_1=0})_{ab'} \equiv \phi_3^{I_{ab'}}. 
\end{equation*}
We omit the subscript ``$mn$" again because they have YM indices $a'b$ or $ab'$ with which 
we can see that these fields are in the ``$mn$" sector. 

The signs of the magnetic fluxes are constrained to yield 
the nonvanishing Yukawa couplings 
$\phi_1^{I_{a'b'}}\phi_2^{I_{b'c}}\phi_3^{I_{ca'}}$ and 
$\phi_2^{I_{a'b}}\phi_1^{I_{bc}}\phi_3^{I_{ca'}}$ 
because the magnetic fluxes cause the chirality projections. 
As a result, the fluxes should satisfy the following conditions on the first torus, 
\begin{equation*}
M_{a'b'}^{(1)}>0,\qquad M_{ab}^{(1)}>0,\qquad 
M_{a'b}^{(1)}<0,\qquad M_{ab'}^{(1)}<0. 
\end{equation*} 
 In the case of vanishing magnetic fluxes, the 4D effective action would be changed 
and is discussed in Appendix A.

On this magnetized background, 
we can derive the 4D effective action for the mixing part $S_{mn}$,  
\begin{eqnarray}
S_{mn} &=& \int~d^{4}x\int~d^4\theta {~\rm Tr~}
\left(\tilde Z^{\bar 22}_{I_{a'b}}\bar\phi_2^{I_{a'b}}e^{-V^{a'}}\phi_2^{I_{a'b}} e^{V^{b}}
+\tilde Z^{\bar 33}_{I_{ab'}}\phi_3^{I_{ab'}} e^{V^a}\bar\phi_3^{I_{ab'}} 
e^{-V^{b'}}\right)\nonumber\\
&&\hspace{0pt}+\int~d^2\theta {~\rm Tr~}
\left[\tilde\lambda_{I_{a'b'}I_{b'c}I_{ca'}}
\phi_1^{I_{a'b'}}\phi_2^{I_{b'c}}\phi_3^{I_{ca'}} 
+\tilde\lambda_{I_{c'a}I_{ab}I_{bc'}}
\phi_2^{I_{c'a}}\phi_1^{I_{ab}}\phi_3^{I_{bc'}} +{\rm h.c.}\right].   
\label{eq:smnaction}
\end{eqnarray}
In this action, the K\"ahler metrics and the holomorphic Yukawa couplings are described as 
\begin{eqnarray*}
\tilde Z^{\bar 22}_{I_{a'b}}&=&2h^{22},\\ 
\tilde Z^{\bar 33}_{I_{ab'}}&=&2h^{33},\\
\tilde\lambda_{I_{a'b'}I_{b'c}I_{ca'}} &=& 
-2g_6\left(e_1e_2e_3\right)^{-1} \tilde\lambda_{a'b',c}^{(1)} \\
\tilde\lambda_{I_{c'a}I_{ab}I_{bc'}} &=& 
2g_{10}\left(e_1e_2e_3\right)^{-1} Q \tilde\lambda_{ab,c'}^{(1)}~~, \\
\end{eqnarray*}
where $Q$ is defined in Eq.~(\ref{eq:afactor}) and $\tilde\lambda_{a'b',c}$ is given by 
\begin{eqnarray*}
\tilde\lambda^{(1)}_{a'b',c} &=& 
{\mathcal N}_{M^{(1)}_{a'b'}}^{-1}{\mathcal N}_{M^{(1)}_{b'c}}{\mathcal N}_{M^{(1)}_{ca'}}
\sum_{m=1}^{M^{(1)}_{a'b'}} 
\delta_{I_{b'c}+I_{ca'}-m M^{(1)}_{b'c},~I_{a'b'}}\\
 && \times 
\exp \left[ \frac{\pi i}{\im\tau_1} 
\left( 
  \frac{\bar\zeta^{(1)}_{a'b'}}{M^{(1)}_{a'b'}}\im\zeta^{(1)}_{a'b'}
+\frac{\bar\zeta^{(1)}_{b'c}}{M^{(1)}_{b'c}}\im\zeta^{(1)}_{b'c}
+\frac{\bar\zeta^{(1)}_{ca'}}{M^{(1)}_{ca'}}\im\zeta^{(1)}_{ca'}
\right) \right] \\
 && \times \vartheta 
\begin{bmatrix}
\frac{
M^{(1)}_{b'c} I_{ca'} - M^{(1)}_{ca'} I_{b'c} + m M^{(1)}_{b'c} M^{(1)}_{ca'}}
{M^{(1)}_{a'b'} M^{(1)}_{b'c} M^{(1)}_{ca'}} 
\\ 0 \end{bmatrix} 
\left( 
 \bar\zeta^{(1)}_{ca'}M^{(1)}_{b'c} - \bar\zeta^{(1)}_{b'c}M^{(1)}_{ca'}, 
-\bar\tau_1 M^{(1)}_{a'b'}M^{(1)}_{b'c}M^{(1)}_{ca'} \right)~,  
\end{eqnarray*}
and $\tilde\lambda_{ab,c'}$ is given by replacing as $(a',b',c)\rightarrow (a,b,c')$ in the above 
expression.

\subsection{Supergravity action and moduli dependence} 
We embed the 4D effective action derived from the mixture of the SYM theories 
into the general form of the conformal SUGRA action (\ref{eq:genesugra}). 
This embedding of the 10D $U(N)$ SYM part $S_n$ is  
the exactly same as is given in the previous section. 

We generalize the discussion given in the previous section to treat 
various dimensional SYM theories. 
Now, we are considering the 6D SYM and the 10D SYM theories, 
and their gauge couplings are described as $g_6$ and $g_{10}$. 
Although we ignored the mass dimension of the gauge coupling in the case with 
only the 10D SYM theory and used the relation $g_{10}=e^{\langle \phi_{10}\rangle}/2$, 
we should renew it more exactly in the generic systems. 
In the $4+2n$ dimensional SYM theories, 
the gauge couplings are determined by the 10D dilaton as 
\begin{equation}
g_{4+2n}=e^{\langle\phi_{10}\rangle/2}{\alpha'}^{n/2},  \label{eq:gaugeconst}
\end{equation}
where $\alpha'$ is a constant parameter and 
it has the mass dimension of $[\rm mass]^{-2}$. 
This parametrization is also supported by the string and D-brane pictures, 
where the parameter $\alpha'$ is equivalent to the square of the string length scale. 
According to this, the definitions of the moduli fields (\ref{eq:moduli}) are also modified 
as 
\begin{equation}
\re \langle S\rangle = e^{-\langle\phi_{10}\rangle}{\alpha'}^{-3}\prod_{i=1}^3\mathcal A^{(i)}, \qquad
\re \langle T_i\rangle = e^{-\langle\phi_{10}\rangle}{\alpha'}^{-1}\mathcal A^{(i)}, \qquad
\langle U_i\rangle = i\bar\tau_i~, \label{eq:modulidef59}
\end{equation}
and the VEV of the 4D dilaton $\phi_4$ is determined as 
\begin{equation*}
e^{-2\langle\phi_4\rangle}
= e^{-2\langle\phi_{10}\rangle}{\alpha'}^{-3}\prod_i\mathcal A^{(i)} 
= \frac1{g_{10}^2}\prod_i\mathcal A^{(i)}. 
\end{equation*}

Before upgrading the parameters to the moduli fields using the above relations, 
the field rescaling should be performed 
to remove some factors to preserve the holomorphicity of the superpotential. 
This operation was also required in single 10D SYM theories. 
In the mixture of the 6D $U(M)$ SYM theory and the 10D $U(N)$ theory, 
we have four types of the Yukawa couplings as follows, 
\begin{eqnarray}
\lambda^{ijk}_{I_{a'b'}I_{b'c'}I_{c'a'}}
\phi_i^{I_{a'b'}}\phi_j^{I_{b'c'}}\phi_k^{I_{c'a'}}
&\qquad& ({\rm three~6D~fields~in~} S_m ),\nonumber\\
\lambda^{ijk}_{\mathcal I_{ab}\mathcal I_{bc}\mathcal I_{ca}}
\phi_i^{\mathcal I_{ab}}\phi_j^{\mathcal I_{bc}}\phi_k^{\mathcal I_{ca}}
&\qquad& ({\rm three~10D~fields~in~} S_n ),\nonumber\\
\lambda_{I_{a'b'}I_{b'c}I_{ca'}}
\phi_1^{I_{a'b'}}\phi_2^{I_{b'c}}\phi_3^{I_{ca'}}
&\qquad& ({\rm mixing~with~a~6D~field~in~} S_{mn} ),\nonumber\\
\lambda_{I_{c'a}\mathcal I_{ab} I_{bc'}}
\phi_2^{I_{c'a}}\phi_1^{\mathcal I_{ab}}\phi_3^{I_{bc'}}
&\qquad& ({\rm mixing~with~a~10D~field~in~} S_{mn} ), \label{eq:4tyukawa}
\end{eqnarray} 
These 4D effective couplings can be decomposed into two parts. 
One is described by the Jacobi-theta function and 
will be holomorphic functions of the moduli fields straightforwardly. 
On the contrast, the other part must be removed to the corresponding K\"ahler metrics 
by the field redefinitions 
because it will contain both the moduli fields and their conjugates simultaneously. 
We focus on the latter part here to determine the rescaling rules 
neglecting trivial numerical factors and the Wilson line parameters. 
The focused part is fortunately universal for the generation structures and shown as 
\begin{eqnarray}
\lambda^{ijk}_{I_{a'b'}I_{b'c'}I_{c'a'}} &\propto& 
e^{3\langle\phi_4\rangle}e^{-K^{(0)}/2} \left(\prod_r 2\pi R_r\right)^{-1} 
g_6\frac{(\im\tau_1)^{1/4}}{\sqrt{\mathcal A_1}},\nonumber\\
\lambda^{ijk}_{\mathcal I_{ab}\mathcal I_{bc}\mathcal I_{ca}} &\propto& 
e^{3\langle\phi_4\rangle}e^{-K^{(0)}/2} \left(\prod_r 2\pi R_r\right)^{-1} 
g_{10}\frac{(\im\tau_1\im\tau_2\im\tau_3)^{1/4}}{\sqrt{\mathcal A_1\mathcal A_2\mathcal A_3}},
\nonumber\\
\lambda_{I_{a'b'}I_{b'c}I_{ca'}} &\propto& 
e^{3\langle\phi_4\rangle}e^{-K^{(0)}/2} \left(\prod_r 2\pi R_r\right)^{-1} 
g_6\frac{(\im\tau_1)^{1/4}}{\sqrt{\mathcal A_1}},\nonumber\\
\lambda_{I_{c'a}\mathcal I_{ab} I_{bc'}} &\propto& 
e^{3\langle\phi_4\rangle}e^{-K^{(0)}/2} \left(\prod_r 2\pi R_r\right)^{-1} 
g_{10}\frac{(\im\tau_1\im\tau_2\im\tau_3)^{1/4}}{\sqrt{\mathcal A_1\mathcal A_2\mathcal A_3}}.  
\label{eq:unholo}
\end{eqnarray} 

The first and the third one are related to the $(a'b')$-fields $\phi_i^{I_{a'b'}}$ originating 
from the 6D SYM theory, 
and the extra dimensional integrals induce the same factors in these two Yukawa couplings. 
And also, the second and the last one, which are related to the $(ab)$-fields originating from 
the 10D $U(N)$ SYM theory, have 
the same form as each other. 
Since there are only three types of the fields to be rescaled, 
the rescaling rules are deterministic. Indeed, they would be uniquely found 
by using the rule for the $(ab)$-fields: 
In the second line, 
the Yukawa coupling of $(ab)$-,$(bc)$- and $(ca)$-fields is shown, 
and it is completely removed in accordance with the rescaling defined in the section~2. 
The forth line expresses the coupling of the $(ab)$-,$(bc')$- and $(c'a)$-fields. 
Since the rescaling factor of the $(ab)$- field is already fixed, 
those of the other two fields are determined naively, 
and then, the rescaling rule for field $\phi_1^{I_{a'b'}}$ is elicited in the third line. 
Finally, the first line determines those for the rest of contents 
$\phi_2^{I_{b'c'}}$ and $\phi_3^{I_{c'a'}}$. 
Note that, the $(a'b')$-sector originates from $\phi_1$, $\phi_2$ and $\phi_3$. 
one originating from $\phi_1$ forms a $\mathcal N=2$ vector multiplet with the 4D vector fields 
and the others form hypermultiplets which can be identified as position moduli fields. 
Thus, it seems sensible that the moduli dependence of 
their K\"ahler metrics are different for $\phi_1^{I_{a'b'}}$ and the other two fields. 

The K\"ahler metrics and the holomorphic Yukawa couplings in the 
generic form of the conformal supergravity can be found 
by the rescaling according to the above discussion. 
Let us start from a review of the 10D $U(M)$ SYM part 
with the renewed moduli definitions (\ref{eq:modulidef59}). 
Although the corresponding factor is shown in the second line of Eq.~(\ref{eq:unholo}), 
its complete form including numerical factors is expressed by 
\begin{eqnarray*}
\lambda^{ijk}_{\mathcal I_{ab}\mathcal I_{bc}\mathcal I_{ca}} &=& 
-\frac{2^{21/4}}3\epsilon ^{ijk}\delta_{\rm i}^i \delta_{\rm j}^j \delta_{\rm k}^k 
e^{3\langle\phi_4\rangle} 
\left(\prod_r 2\pi R_r\right)^{-1} \left(\prod_{r'} \re \langle T_{r'}\rangle\right)^{1/2} 
\left(\prod_{r''} \re \langle U_{r''}\rangle \right)^{3/4} \\
&&\times
\left|\frac{M_{ab}^{(2)}M_{ab}^{(3)}}{M_{ab}^{(1)}}\right|^{1/4}
\left|\frac{M_{bc}^{(1)}M_{bc}^{(3)}}{M_{bc}^{(2)}}\right|^{1/4}
\left|\frac{M_{ca}^{(1)}M_{ca}^{(2)}}{M_{ca}^{(3)}}\right| ^{1/4}\times e^H\times \vartheta, 
\end{eqnarray*}
where the exponential factor of the Wilson lines $e^H$ corresponds to 
the the second line of Eq.~(\ref{eq:yukawaxxx}) and 
the holomorphic part given by the Jacobi-theta function is represented 
by the last factor $\vartheta$. 
The following field rescaling recovers the K\"ahler metric and the holomorphic Yukawa coupling 
of the form obtained in section~2, 
\begin{eqnarray}
\phi_i^{\mathcal I_{ab}} &\rightarrow& \alpha_i^{ab} \phi_i^{\mathcal I_{ab}},
\label{eq:resca10d} 
\end{eqnarray}
where 
\begin{eqnarray}
 \alpha_i^{\mathcal I_{ab}} &=& 2^{-7/4} e^{-\langle\phi_4\rangle} 
\frac{2\pi R_i}{\sqrt{\re \langle T_i\rangle }} 
\left(\prod_{r} \re \langle U_{r}\rangle \right)^{-1/4} \nonumber\\
&&\qquad \times {\rm exp}
\left[-\sum_r\frac{\pi i}{\im\tau_r}\frac{\bar\zeta_{ab}^{(r)}}{M_{ab}^{(r)}}\im\zeta_{ab}^{(r)}\right] 
\left(\frac{|M_{ab}^{(i)}|}{\prod_{j\neq i}|M_{ab}^{(j)}|}\right)^{1/4}. 
\label{eq:resca10d2} 
\end{eqnarray} 
The gauge kinetic function shown in Eq.~(\ref{eq:kineticg}) is also recovered 
using the gauge coupling (\ref{eq:gaugeconst}) and moduli definition (\ref{eq:modulidef59}). 

Next we lead to the K\"ahler metrics of $(ab')$- and $(a'b)$-fields 
contained in the mixing part $S_{mn}$. 
From the second and the fourth lines of Eq.~(\ref{eq:unholo}), 
it is inferred that the rescaling factors of these fields are equivalent to $(ab)$-sector 
up to numerical factors and the exponential factors of the Wilson lines. 
Indeed, 
when the localization described by the delta function (\ref{eq:tdelta}) is not shifted (
We will also discuss in another case of shifted quasi-localizations later. ), 
the un-holomorphic part of $\lambda_{I_{c'a}\mathcal I_{ab}I_{bc'}}$ is entirely removed 
by the rescaling  (\ref{eq:resca10d}) and 
\begin{eqnarray}
\phi_2^{I_{c'a}} &\rightarrow& \alpha_2^{c'a} \phi_2^{I_{c'a}}, \qquad 
\phi_3^{I_{bc'}} \rightarrow \alpha_3^{bc'} \phi_3^{I_{bc'}}, 
\label{eq:rescabifun}
\end{eqnarray}
where 
\begin{eqnarray*}
\alpha_2^{c'a} &=& 2^{-7/4} e^{-\langle\phi_4\rangle} 
\frac{2\pi R_2}{\sqrt{\re \langle T_2\rangle }} 
\left(\prod_{r} \re \langle U_{r}\rangle \right)^{-1/4}\\
&&\qquad\times {\rm exp}
\left[-\frac{\pi i}{\im\tau_1}\frac{\bar\zeta_{c'a}^{(1)}}{M_{c'a}^{(1)}}\im\zeta_{c'a}^{(1)}\right] 
 |M_{c'a}^{(1)}|^{-1/4}, \\
\alpha_3^{bc'} &=& 2^{-7/4} e^{-\langle\phi_4\rangle} 
\frac{2\pi R_3}{\sqrt{\re \langle T_3\rangle }} 
\left(\prod_{r} \re \langle U_{r}\rangle \right)^{-1/4} \\
&&\qquad 
\times {\rm exp}
\left[-\frac{\pi i}{\im\tau_1}\frac{\bar\zeta_{bc'}^{(1)}}{M_{bc'}^{(1)}}\im\zeta_{bc'}^{(1)}\right] 
 |M_{bc'}^{(1)}|^{-1/4}.   
\end{eqnarray*} 
After these rescalings, the relevant holomorphic Yukawa couplings are found as 
\begin{equation*}
\lambda_{I_{c'a}\mathcal I_{ab}I_{bc'}} = 
\lambda^{(1)}_{ab,c'}\times \left(
\prod_{r=2,3}
\vartheta\begin{bmatrix}
I_{ab}^{(r)}/M_{ab}^{(r)}\\0\end{bmatrix}\left(\zeta_{ab}^{(r)}, iM_{ab}^{(r)}\bar U_r\right)
\right), 
\end{equation*} 
where $\lambda^{(1)}_{ab,c}$ is found in Eq.~(\ref{eq:finalyukaholo}) 
by the replacing $c\rightarrow c'$, and then, the K\"ahler metrics 
of two types of the bifundamental fields are obtained as 
\begin{eqnarray*}
Z_{I_{c'a}^{\bar 22}} &=& 
\frac1{2^{5/2}} \left(\frac{T_2+\bar T_2}2\right)^{-1} 
\left(\prod_{r} \frac{U_r+\bar U_r}2 \right)^{-1/2} \\
&&\qquad 
\times{\rm exp}\left[-\frac{4\pi}{U_1+\bar U_{\bar 1}}
\frac{\left(\im\zeta_{c'a}^{(1)}\right)^2}{M_{ca'}^{(1)}}\right]
|M_{c'a}^{(1)}|^{-1/2},\\
Z_{ I_{bc'}^{\bar 33}} &=& 
\frac1{2^{5/2}} \left(\frac{T_3+\bar T_3}2\right)^{-1} 
\left(\prod_{r} \frac{U_r+\bar U_r}2 \right)^{-1/2} \\
&&\qquad 
\times{\rm exp}\left[-\frac{4\pi}{U_1+\bar U_{\bar 1}}
\frac{\left(\im\zeta_{bc'}^{(1)}\right)^2}{M_{bc'}^{(1)}}\right]
|M_{bc'}^{(1)}|^{-1/2}. 
\end{eqnarray*}


The rescaling factor of $\phi_1^{I_{a'b'}}$ 
is derived from the Yukawa couplings shown in the third line of 
Eq~(\ref{eq:unholo}). 
As the result, it is found as 
\begin{eqnarray*}
\alpha_1^{a'b'} &=& 2^{-5/4} e^{-\langle\phi_4\rangle} 
\frac{2\pi R_1}{\sqrt{\re \langle S\rangle }} 
\left( \re \langle U_{1}\rangle \right)^{-1/4} \\
&&\qquad\times {\rm exp}
\left[-\frac{\pi i}{\im\tau_1}\frac{\bar\zeta_{a'b'}^{(1)}}{M_{a'b'}^{(1)}}\im\zeta_{a'b'}^{(1)}\right] 
 |M_{a'b'}^{(1)}|^{1/4}.  
\end{eqnarray*} 
This yields the K\"ahler metric of the form 
\begin{eqnarray*}
Z_{I_{a'b'}^{\bar 11}} &=& 
\frac1{2^{3/2}} \left(\frac{S+\bar S}2\right)^{-1} 
\left(\frac{U_1+\bar U_1}2 \right)^{-1/2} \\
&&\qquad 
\times{\rm exp}\left[-\frac{4\pi}{U_1+\bar U_{\bar 1}}
\frac{\left(\im\zeta_{a'b'}^{(1)}\right)^2}{M_{a'b'}^{(1)}}\right]
|M_{a'b'}^{(1)}|^{1/2},\\
\end{eqnarray*} 
and the holomorphic Yukawa coupling is simply given by 
\begin{equation*}
\lambda_{I_{a'b'}I_{b'c}I_{ca'}} =-\lambda^{(1)}_{a'b',c}~~.
\end{equation*}

Finally, the rest of the rescaling factors are automatically determined 
in the first line of  Eq.~(\ref{eq:unholo}) as 
\begin{equation*}
\phi_j^{I_{b'c'}}\rightarrow\alpha_j^{b'c'}\phi_j^{I_{b'c'}}\qquad{\rm for~}j=2,3,
\end{equation*}
where 
\begin{eqnarray*} 
\alpha_j^{b'c'} &=& 2^{-7/4} e^{-\langle\phi_4\rangle} 
\frac{2\pi R_j}{\sqrt{\re \langle T_j\rangle }} 
\left(\prod_{r} \re \langle U_{r}\rangle \right)^{-1/4}  \\
&&\qquad\times {\rm exp}
\left[-\frac{\pi i}{\im\tau_1}\frac{\bar\zeta_{b'c'}^{(1)}}{M_{b'c'}^{(1)}}\im\zeta_{b'c'}^{(1)}\right] 
 |M_{b'c'}^{(1)}|^{-1/4} .
\end{eqnarray*} 
Their K\"ahler metrics and holomorphic Yukawa couplings are found as follows, 
\begin{eqnarray*}
Z_{\mathcal I_{b'c'}^{\bar jj}} &=& 
\frac1{2^{5/2}} \left(\frac{T_j+\bar T_j}2\right)^{-1} 
\left(\prod_{r} \frac{U_r+\bar U_r}2 \right)^{-1/2} \\
&&\qquad 
\times{\rm exp}\left[-\frac{4\pi}{U_1+\bar U_{\bar 1}}
\frac{\left(\im\zeta_{b'c'}^{(1)}\right)^2}{M_{b'c'}^{(1)}}\right]
|M_{b'c'}^{(1)}|^{-1/2}\\
\end{eqnarray*}
for $j=2,3$, and 
\begin{equation*}
\lambda_{I^{(1)}_{a'b'}I^{(1)}_{b'c'}I^{(1)}_{c'a'}}^{ijk} =-\frac13\epsilon^{\rm ijk}
\delta_{\rm i}^i \delta_{\rm j}^j \delta_{\rm k}^k\lambda^{(1)}_{a'b',c'}~~.
\end{equation*}

The gauge kinetic functions of the $U(M)$ subgroups are given by  
\begin{equation*}
f_{a'} = T_1, 
\end{equation*}
which is different from those derived from the 10D $U(N)$ SYM theory. 
This is one of significant features of the mixed higher-dimensional SYM systems. 
This is also consistent with the interpretation in a D-brane picture. 

In the rest of this section, we discuss another case in which 
the point-like quasi-localizations of $(a,b')$- and $(a',b)$- sectors are shifted by VEVs of 
the position moduli. 
This effect appear in the factor $Q$ of $(c'a)-(ab)-(bc')$ coupling, 
which is defined in Eq.~(\ref{eq:afactor}). 
Considering the point like-localizations shifted from $z_s=0$ by $\chi_s$ (s=2,3), 
the factor $Q$ is given by 
\begin{equation*}
Q =\prod_{s=2,3}\left\{ \mathcal N_{M^{(s)}_{ab}}e^{\frac{\pi i}{\im\tau_s}M^{(s)}_{ab}
\left(\bar\chi_s+\frac{\bar\zeta_{ab}^{(s)}}{M^{(s)}_{ab}}\right)
\im\left(\chi_s+\frac{\zeta_{ab}^{(s)}}{M^{(s)}_{ab}}\right)} 
\times \vartheta\right\}. 
\end{equation*}
When $\chi_s$ is vanishing, the rescaling of $\phi_1^{\mathcal I_{ab}}$ 
defined in Eqs.~(\ref{eq:resca10d}) and (\ref{eq:resca10d2}) 
consistently removes the above exponential factor. 
We extract additional contributions induced by nonvanishing $\chi_s$ from the above 
equation as 
\begin{equation*}
Q\propto\exp\left[\sum_{s=2,3}\frac{\pi i}{\im\tau_s}
\left(M^{(s)}_{ab}\bar\chi_s\im\chi_s+\chi_s\im\zeta_{ab}^{(s)}+\bar\zeta_{ab}^{(s)}\im\chi_s
\right)\right]
\end{equation*}
This will be absorbed by the rescaling of $(ab)$-, $(bc')$- and $(c'a)$-sectors. 
When we consider a modification of the rescaling rule for  $(bc')$- and $(c'a)$-sectors 
to remove this factor, 
those for $(a'b')$-sector (6D fields) must also be modified for the holomorphicity of 
Yukawa couplings. 
As a result, the shift parameter $\chi_s$ appear in K\"ahler metrics of the 6D fields, 
even though this shift is caused in four-dimensional extra compact space which is not related to 
the 6D field theory. This is a bizarre consequence and we should consider another way. 
Thus, this additional factor would be absorbed by only $(ab)$-sector, 
and then, $\phi_1^{\mathcal I_{ab}}$ is further rescaled as 
\begin{eqnarray}
\phi_1^{\mathcal I_{ab}}&\rightarrow& \tilde\alpha_1^{\mathcal I_{ab}}\phi_1^{\mathcal I_{ab}}
\nonumber\\
\tilde\alpha_1^{\mathcal I_{ab}} &=& {\rm exp}
\left[\sum_{s=2,3}-\frac{\pi i}{\im\tau_s}\left(M^{(s)}_{ab}\bar\chi_s\im\chi_s+\chi_s\im\zeta_{ab}^{(s)}
+\bar\zeta_{ab}^{(s)}\im\chi_s
\right)\right].\label{eq:furresca} 
\end{eqnarray}
As the result, the K\"ahler metric of $\phi_1^{\mathcal I_{ab}}$ is found as 
\begin{eqnarray}
Z_{\mathcal I_{ab}^{\bar ij}} &=& \delta^{\bar ij}
\left(\frac{T_j+\bar T_{\bar j}}{2}\right)^{-1}
\left(\prod_{r=1}^3\frac{U_r+\bar U_{\bar r}}{2}\right)^{-1/2}
\nonumber\\
&&\qquad
\times\frac1{2^{5/2}}
\left(\frac{|M_{ab}^{(j)}|}{\prod_{r\neq j}|M_{ab}^{(r)}|}
\right)^{1/2}
{\rm exp}\left[-\sum_{r=1}^3\frac{4\pi}{U_r+\bar U_{\bar r}}
\frac{\left(\im\zeta_{ab}^{(r)}\right)^2}{M_{ab}^{(r)}}\right], 
\nonumber\\
&&\qquad\qquad
\times
{\rm exp}\left[-\sum_{s=2,3}^3\frac{4\pi }{U_s+\bar U_{\bar s}}
\left(M^{(s)}_{ab}\left(\im\chi_s\right)^2+2\im\chi_s\im\zeta_{ab}^{(s)}
\right)\right], \label{eq:finalzii}
\end{eqnarray}
where $i=\bar i=1$ and the last line represents the additional contribution. 
This rescaling of $\phi_1^{\mathcal I_{ab}}$ induce 
the additional factor $\tilde\alpha_i^{\mathcal I_{ab}}$ 
in another Yukawa coupling 
$\phi_1^{\mathcal I_{ab}}\phi_j^{\mathcal I_{bc}}\phi_k^{\mathcal I_{ca}}$
($j,k=2,3$) 
shown in the second line of Eq.~(\ref{eq:4tyukawa}), 
but rescalings of $\phi_j^{\mathcal I_{bc}}$ and $\phi_k^{\mathcal I_{ca}}$ 
can naturally absorb this factor. 
This is because that 
the additional factor (\ref{eq:furresca} ) is rewritten as 
\begin{eqnarray*}
\tilde\alpha_i^{\mathcal I_{ab}} &=& {\rm exp}
\left[\sum_{s=2,3}\frac{\pi i}{\im\tau_s}\left(
M^{(s)}_{bc}\bar\chi_s\im\chi_s+\chi_s\im\zeta_{bc}^{(s)}
+\bar\zeta_{bc}^{(s)}\im\chi_s \right.\right.\\
&&\qquad\qquad\qquad\quad\left.\left.+ M^{(s)}_{ca}\bar\chi_s\im\chi_s+\chi_s\im\zeta_{ca}^{(s)}
+\bar\zeta_{ca}^{(s)}\im\chi_s
\right)\right], 
\end{eqnarray*}
where we use 
$M^{(s)}_{ab}+M^{(s)}_{bc}+M^{(s)}_{ca}=0 (\zeta^{(s)}_{ab}+\zeta^{(s)}_{bc}+\zeta^{(s)}_{ca}=0)$. 
This is removed by further rescalings of $\phi_j^{\mathcal I_{bc}}$ and $\phi_k^{\mathcal I_{ca}}$ 
as follows, 
\begin{eqnarray*}
\phi_j^{\mathcal I_{bc}}&\rightarrow& \tilde\alpha_j^{\mathcal I_{bc}}\phi_j^{\mathcal I_{bc}}, 
\qquad 
\phi_k^{\mathcal I_{ca}}\rightarrow \tilde\alpha_k^{\mathcal I_{ca}}\phi_k^{\mathcal I_{ca}}, \\
\tilde\alpha_j^{\mathcal I_{bc}} &=& {\rm exp}
\left[\sum_{s=2,3}-\frac{\pi i}{\im\tau_s}\left(M^{(s)}_{bc}\bar\chi_s\im\chi_s+\chi_s\im\zeta_{bc}^{(s)}
+\bar\zeta_{bc}^{(s)}\im\chi_s
\right)\right],\\
\tilde\alpha_k^{\mathcal I_{ca}} &=& {\rm exp}
\left[\sum_{s=2,3}-\frac{\pi i}{\im\tau_s}\left(M^{(s)}_{ca}\bar\chi_s\im\chi_s+\chi_s\im\zeta_{ca}^{(s)}
+\bar\zeta_{ca}^{(s)}\im\chi_s
\right)\right].  
\end{eqnarray*}
These have the same form as Eq.~(\ref{eq:furresca}). 
And also, the expression (\ref{eq:finalzii}), 
which gives the K\"ahler metric of $\phi_1^{\mathcal I_{ab}}$ for $i=\bar i=1$, 
can describe those of the other two fields for $i=\bar i=2,3$. 
We have obtained the general form of the 4D effective action 
which is valid even when the positions of the 6D fields on the two tori are 
shifted by the nonvanishing VEVs of the position moduli. 
A variety of 6D and 10D SYM systems is obtained with multiple 6D SYM theories 
distinguished by their localized points. 
Furthermore, the most generic SYM system can also be also constructed 
in the similar way which is demonstrated in this section. 

In the next section, we show a mixed system consisting of 
4D SYM theories and 8D SYM theories as another example. 

\section{4D and 8D SYM theories and their mixtures}
\label{sec:4d8d}
Although any of SYM mixtures basically can be derived in the same manner, 
we give another specific system with 4D and 7D SYM theories. 
These SYM theories can be expected to appear as low-energy effective field theories 
of mixed configurations of D3- and D7-branes. 
It is known that the D3-D7 brane systems are related to the D9-D5 brane systems by T-duality. 
Indeed, SYM systems which might describe these two D-brane systems 
can be derived from a single 10D SYM theory 
with the same configuration of the infinite magnetic fluxes (\ref{eq:fluxinft}) 
by performing two different ways of dimensional reduction. 

\subsection{Superfield description of the 4D and 8D SYM theories}
We derive a superfield description of mixture 
of a 4D $U(N)$ SYM theory localized at a point of the extra dimensions and 
an 8D $U(M)$ SYM theory compactified on two tori, $M^4\times(T^2)_2\times(T^2)_3$ 
from the 10D $U(M+N)$ SYM theory with the infinite magnetic fluxes (\ref{eq:fluxinft}). 
The remaining zero-modes are also equivalent to those of the previous model 
but they are interpreted differently. 
The first-block entries are assigned to the 8D $U(M)$ SYM theory and 
the last ones to the 4D $U(N)$ SYM theory. 
In these theories, 
$\phi_2^n$ and $\phi_3^n$ are identified as the position moduli of the $U(N)$ SYM theory 
on the two tori. 
$\phi_1^m$ and $\phi_1^n$ can also be seen as the position moduli 
of the irrelevant torus $(T^2)_1$ 
where no field of this system lives. 
For simplicity, the VEVs of $\phi_1^m$ and $\phi_1^n$ are set to vanish in the following.

The effective action obtained by the partial dimensional reduction 
is given by the following three parts, 
\begin{equation*}
S=S_m+S_n+S_{mn},
\end{equation*} 
where $S_m$ corresponds to the 8D $U(M)$ SYM theory 
compactified on the second and the third tori, 
$S_n$ to the 4D $U(N)$ SYM theory and 
the last part $S_{mn}$ to the mixings of the two theories which 
contains $\phi_2^{mn}$ and $\phi_3^{mn}$. 
These are easily calculated in a similar way to the previous section: 
The 8D $U(M)$ SYM theory is obtained by carrying out the integration with respect to 
coordinates $z_1$ and $\bar z_{\bar1}$, and it is found as 
\begin{equation}
S_m=\frac{1}{g_8^2}
\int d^8X\sqrt{-G_8}\int d^4\theta \mathcal K_m +\left\{
\int d^2\theta\left(\frac1{4}\mathcal W_m^\alpha
{\mathcal W_\alpha}_m
+\mathcal W_m \right)+{\rm h.c.}\right\},\nonumber
\end{equation}
where the three functions $\mathcal K_m$, $\mathcal W_m$ and 
$\mathcal W_m^\alpha$  are given by 
\begin{eqnarray}
\mathcal K_m &=& 2{\rm Tr}\left[
h^{22}\left(\left(\sqrt2\bar\partial_2+\bar\phi_2^m\right)e^{-V^m}\right)
\left(-\sqrt2\partial_2+\phi_2^m\right)e^{V^m}\right.\nonumber\\
& & \hspace{20pt}+
h^{33}\left(\left(\sqrt2\bar\partial_3+\bar\phi_3^m\right)e^{-V^m}\right)
\left(-\sqrt2\partial_3+\phi_3^m\right)e^{V^m}\nonumber\\
& & \hspace{20pt}+h^{22}\bar\partial_2e^{-V^m}\partial_2e^{V^m}
+h^{33}\bar\partial_3e^{-V^m}\partial_3e^{V^m}\nonumber\\
& & \hspace{20pt}\left.+h^{11}\bar\phi_1^m e^{-V^m}\phi_1^m e^{V^m}
+\mathcal K'_{\rm WZW}\right],\nonumber\\
\mathcal W_m &=& 2\sqrt2 \left(e_1e_2e_3\right)^{-1}
\left(\phi_3^m\partial_1\phi_2^m+\phi_1^m\partial_2\phi_3^m
-\frac{1}{\sqrt2}\phi_3^m\left[\phi_1^m, \phi_2^m\right]\right),\nonumber\\
\mathcal {W_\alpha}_m &=& -\frac14\bar D\bar De^{-V^m}D_\alpha e^{V^m}.  
\nonumber
\end{eqnarray}
The 4D $U(N)$ part is given by 
\begin{equation}
S_n=\frac{1}{g_4^2}
\int d^4X\sqrt{-G_4}\int d^4\theta \mathcal K_n +\left\{
\int d^2\theta\left(\frac14\mathcal W_n^\alpha
{\mathcal W_\alpha}_n
+\mathcal W_n \right)+{\rm h.c.}\right\},\nonumber
\end{equation}
where the three functions $\mathcal K_n$, $\mathcal W_n$ and 
$\mathcal W_n^\alpha$ are given by 
\begin{eqnarray}
\mathcal K_n &=& 2{\rm Tr}\left[h^{j\bar i}\bar\phi_{\bar i}^n e^{-V^n}\phi_j^n e^{V^n}
+\mathcal K'_{\rm WZW}\right],\nonumber\\
\mathcal W_n &=& -\frac23\epsilon^{\rm ijk} e_{\rm i}^{~i}e_{\rm j}^{~j}e_{\rm k}^{~k}
\phi_i^n\phi_j^n, \phi_k^n,\nonumber\\
\mathcal {W_\alpha}_n &=& -\frac14\bar D\bar De^{-V^n}D_\alpha e^{V^n}.  
\nonumber
\end{eqnarray} 

After the integration of the well-defined delta functions induced by the infinite magnetic fluxes 
with respect to the torus coordinates, 
the last mixing part $S_{mn}$ is described by 
\begin{eqnarray}
S_{mn} &=& \int~d^{4}X\sqrt{-G_4}\int~d^4\theta {~\rm Tr~}
\left(2h^{22}\bar\phi_2^{mn}e^{-V^m}\phi_2^{mn} e^{V^n}
+2h^{33}\phi_3^{mn}e^{V^m}\bar\phi_3^{mn}e^{-V^n}\right)\nonumber\\
&&\hspace{10pt}+2\sqrt2\left(e_1e_2e_3\right)^{-1} \int~d^2\theta {~\rm Tr~}
\left[\phi_3^{mn}\left(-\frac {\tilde Q}{\sqrt2}\phi_1^m\phi_2^{mn}
+\frac 1{\sqrt2}\phi_2^{mn}\phi_1^n\right) +{\rm h.c.}\right],  
\end{eqnarray}
where the factor $\tilde Q$ is given by the integrals on the two tori, 
\begin{eqnarray}
\tilde Q&=&\prod_{s=2,3}\int dz^sd\bar z^{\bar s} 
\left\{(f_1^{(s)})^{m}(z_s)\times \delta_{T^2}(z_s+\tilde\zeta^{(s)})\right\}, \nonumber\\
&=&\prod_{s=2,3} (f_1^{(s)})^{m} (\tilde\zeta^{(s)}), 
\end{eqnarray}
with $\tilde\zeta^{(s)}\equiv\zeta_{MN}^{(s)}/H$. 

\subsection{4D effective action on magnetized backgrounds}
We derive the 4D effective action from the mixture of the 4D $U(N)$ SYM theory 
and 8D $U(M)$ SYM theory compactified on magnetized tori. 
This is given by the following configuration of magnetic fluxes, instead of Eq.~(\ref{eq:fluxinft})
\begin{eqnarray}
M^{(1)}&=&\begin{pmatrix}
0 &0\\
0&0
\end{pmatrix},\nonumber\\
M^{(2)}=\begin{pmatrix}
M^{(2)}_m+H\times{\bm 1}_{M}&0\\
0&0
\end{pmatrix}, & &\qquad
M^{(3)}=\begin{pmatrix}
M^{(3)}_m-H\times{\bm 1}_{M}&0\\
0&0
\end{pmatrix},
\end{eqnarray} 
where the finite fluxes of the 8D $U(M)$ SYM theory $M^{(2)}_m$ and $M^{(3)}_m$ are 
$(M\times M)$ matrices and they can lead to 
a gauge symmetry breaking $U(M)\rightarrow \prod_a U(M_a)$. 
Note that, in this configuration, the infinite fluxes $H$ and $-H$ can be moved 
to the last-block entries without any physical changes 
(as long as we are studying SYM theories).

In assumption of the gauge symmetry breaking $U(M)\rightarrow \prod_a U(M_a)$, 
bifundamental fields $\phi_2^{mn}$ and $\phi_3^{mn}$ appearing in $S_{mn}$ 
are replaced by $\phi_2^{an}$ and $\phi_3^{an}$, which are 
bifundamental representations $(M_a, \bar N)$ and $(\bar M_a, N)$ of 
$U(M_a)\times U(N)$, respectively. 
We can concentrate on the 8D $U(M)$ SYM theory $S_m$
to derive the 4D effective theory of this system 
because the extra dimensional integrations have already been carried out in the other parts.

In assumption of the gauge symmetry breaking $U(M)\rightarrow \prod_a U(M_a)$, 
the 4D effective action of the 8D $U(M)$ SYM $S_{m}$ is given by 
\begin{eqnarray}
S_m=\int d^4 x \left[
\int d^4\theta {\mathcal K_{\rm eff}}+\left\{
\int d^2\theta\left(\frac1{4g_a^2}\mathcal W^{a,\alpha}\mathcal W^a_\alpha 
+\mathcal W_{\rm eff}\right)+ {\rm h.c.}
\right\}\right] ,
\end{eqnarray}
where the functions $\mathcal K_{\rm eff}$, $\mathcal W_{\rm eff}$ and 
$\mathcal W^a_\alpha$ have the following form, 
\begin{eqnarray*}
\mathcal K_{\rm eff} &=& \sum_{i,j}\sum_{a,b}\sum_{\mathcal I_{ab}} 
\tilde Z_{\mathcal I_{ab}}^{\bar i j} {\rm Tr} \left[ 
\bar\phi_{\bar i}^{\mathcal I_{ab}} e^{-V^a} \phi_j^{\mathcal I_{ab}} e^{V^a} \right], \\
\mathcal W_{\rm eff} &=& \sum_{i,j,k}\sum_{a,b,c}
\sum_{\mathcal I_{ab},\mathcal I_{bc},\mathcal I_{ca}}
\tilde\lambda^{ijk}_{\mathcal I_{ab}\mathcal I_{bc}\mathcal I_{ca}}{\rm Tr} 
\left[\phi_i^{\mathcal I_{ab}}\phi_j^{\mathcal I_{bc}}\phi_k^{\mathcal I_{ca}}\right],\\
\mathcal W_\alpha &=& -\frac14\bar D\bar D e^{-V^a}D_\alpha e^{V^a}.\qquad 
g_a = g_8 \left(\mathcal A^{(2)}\mathcal A^{(3)}\right)^{-1/2}, 
\end{eqnarray*}
with $\mathcal I_{ab}=(I_{ab}^{(2)},I_{ab}^{(3)})$, 
and K\"ahler metric $\tilde Z_{\mathcal I_{ab}}^{\bar i j}$ 
and holomorphic Yukawa coupling 
$\tilde\lambda^{ijk}_{\mathcal I_{ab}\mathcal I_{bc}\mathcal I_{ca}}$ 
are given by
\begin{eqnarray*}
\tilde Z_{\mathcal I_{ab}}^{\bar i j} &=& 2h^{\bar ij}\\
\tilde\lambda^{ijk}_{\mathcal I_{ab}\mathcal I_{bc}\mathcal I_{ca}} &=& 
-\frac{2g_8}3\epsilon^{\rm ijk}e_{\rm i}^{~i}e_{\rm j}^{~j}e_{\rm k}^{~k} \prod_{r=2}^3 
\tilde\lambda^{(r)}_{I^{(r)}_{ab} I^{(r)}_{bc} I^{(r)}_{ca}} . 
\end{eqnarray*}
The factor $\tilde\lambda^{(r)}_{I^{(r)}_{ab} I^{(r)}_{bc} I^{(r)}_{ca}}$ is defined in 
Eqs.~(\ref{eq:3ways}) and (\ref{eq:yukawaxxx}) and 
these are valid when $M^{(r)}_{ab}M^{(r)}_{bc}M^{(r)}_{ca}>0$.

\subsection{Supergravity action and moduli dependence}
At this final step, we embed the 4D effective action into the generic form of 
$\mathcal N=1$ conformal supergravity. 
A D3/D7 brane system, which is a motivation of this section, 
is T-dual to a D5/D9 system as we mentioned, and indeed, 
a part of the T-dual picture have been seen in our study of SYM systems. 
According to the T-duality, 
the moduli definitions (\ref{eq:modulidef59}) should also be replaced 
in the 4D- and 8D-SYM systems by 
\begin{equation}
\re \langle S\rangle = e^{-\langle\phi_{10}\rangle}, \qquad
\re \langle T_i\rangle = e^{-\langle\phi_{10}\rangle}{\alpha'}^{-2}\mathcal A^{(j)}\mathcal A^{(k)}, 
\qquad \langle U_i\rangle = i\bar\tau_i~, \label{eq:modulidef37}
\end{equation}
where $i\neq j\neq k\neq i$. 
We can see from studying the gauge kinetic functions that these new identification 
of the moduli VEVs is plausible in our system. 
The two gauge kinetic functions of the 4D effective field theories derived in this section 
are 
\begin{equation*}
\re f_{\rm 4D}=\frac1{g_4^2},\qquad 
\re f_a=\frac1{g_8^2}\mathcal A^{(2)}A^{(3)}. 
\end{equation*}
The parameters in these functions are upgraded to the moduli field 
in accordance with Eqs.~(\ref{eq:gaugeconst}) and (\ref{eq:modulidef37}), 
and that leads to 
\begin{equation*}
f_{\rm 4D}=S,\qquad f_a=T_1. 
\end{equation*}
These results are consistent with the D3/D7 brane picture. 

The field rescalings are also required in this system 
before upgrading the parameters to the moduli fields. 
We first determine the simplest rescaling rule for the fields of 4D $U(N)$ SYM theory $\phi_i^n$ 
which has no generation structure because they are defined in the 4D spacetime 
from the beginning. 
Those for the other fields are uniquely found for the holomorphicity 
of four types of Yukawa couplings. 
The complete form of the Yukawa coupling $\lambda^{ijk}\phi_i^n\phi_j^n\phi_k^n$ is given by 
\begin{equation*}
\lambda^{ijk}=-\frac{2^{9/2}}3\epsilon^{\rm ijk}\delta_{\rm i}^{~i}\delta_{\rm j}^{~j}\delta_{\rm k}^{~k} 
e^{3\langle\phi_4\rangle} 
\left(\prod_r 2\pi R_r\right)^{-1}
\left(\prod_{r'}\re\langle T_{r'}\rangle\right)^{1/2} 
\left(\prod_{r'}\re\langle U_{r''}\rangle\right)^{1/2}. 
\end{equation*}
These are removed by the field rescaling 
\begin{equation*}
\phi_i^n\rightarrow \alpha_i^n\phi_i^n, 
\end{equation*}
where 
\begin{equation*}
\alpha_i^n=2^{-3/2}e^{-\langle\phi_4\rangle} 
\frac{2\pi R_i}{\sqrt{\re\langle T_{i}\rangle}} 
\left(\prod_{r}\re\langle U_{r}\rangle\right)^{-1/6} . 
\end{equation*}
As the result, the K\"ahler metric of this field and the holomorphic Yukawa couplings 
are found as 
\begin{eqnarray*}
Z^n_{i\bar i} &=& \frac14 \left(\frac{T_i+\bar T_i}2\right)^{-1} 
\left(\prod_{r}\frac{U_r+\bar U_r}2\right)^{-1/3}, \\
\lambda^{ijk} &=& -\frac{1}3\epsilon^{\rm ijk} 
\delta_{\rm i}^{~i}\delta_{\rm j}^{~j}\delta_{\rm k}^{~k}.  
\end{eqnarray*} 
This leads to the following results for the other fields: 
\begin{eqnarray*}
\phi_i^{an}&\rightarrow& \alpha_i^{an}\phi_i^{an}, \\
\phi_1^{\mathcal I_{ab}}&\rightarrow& \alpha_i^{ab}\phi_1^{\mathcal I_{ab}}, \\
\phi_j^{\mathcal I_{ab}}&\rightarrow& \alpha_j^{ab}\phi_j^{\mathcal I_{ab}},\qquad j\neq1,
\end{eqnarray*}
where 
\begin{eqnarray*}
\alpha_i^{an}  &=& 
2^{-3/2}e^{-\langle\phi_4\rangle} 
\frac{2\pi R_i}{\sqrt{\re\langle T_{i}\rangle}} 
\left(\prod_{r}\re\langle U_{r}\rangle\right)^{-1/6}, \\
\alpha_1^{ab} &=& 
2^{-2}e^{-\langle\phi_4\rangle} 
\frac{2\pi R_i}{\sqrt{\re\langle S\rangle}} 
\left(\re\langle U_{1}\rangle\right)^{-1/6}
\left(\re\langle U_{2}\rangle\right)^{-5/12}\left(\re\langle U_{3}\rangle\right)^{-5/12}\\
&&\qquad \times {\rm exp}
\left[-\sum_{k\neq1}\frac{\pi i}{\im\tau_k}
\frac{\bar\zeta_{ab}^{(k)}}{M_{ab}^{(k)}}\im\zeta_{ab}^{(k)}\right] 
\left(\prod_{k\neq 1}|M_{ab}^{(k)}|\right)^{-1/4} ,\\
\alpha_j^{ab} &=&
2^{-3/2}e^{-\langle\phi_4\rangle} 
\frac{2\pi R_j}{\sqrt{\re\langle T_{j}\rangle}} 
\left(\prod_{r}\re\langle U_{r}\rangle\right)^{-1/6}\\
&&\qquad \times {\rm exp}
\left[-\sum_{k\neq1}\frac{\pi i}{\im\tau_k}
\frac{\bar\zeta_{ab}^{(k)}}{M_{ab}^{(k)}}\im\zeta_{ab}^{(k)}\right] 
\left(\frac{|M_{ab}^{(j)}|}{|M_{ab}^{(s)}|_{j\neq s\neq1}}\right)^{1/4}.   
\end{eqnarray*}
Recall that $\phi_1^n$ is the $U(N)$ adjoint representation, 
$\phi_i^{\mathcal I_{ab}}$ is the bifundamental representation $(M_a,\bar M_b)$, 
and $\phi_2^{an}$ and $\phi_3^{an}$ are $(M_a, \bar N)$ and $(\bar M_a, N)$, respectively. 
The K\"ahler metrics for these fields are given by 
\begin{eqnarray*}
Z^{an}_{j\bar j} &=& 
\frac14 \left(\frac{T_j+\bar T_j}2\right)^{-1} 
\left(\prod_{r}\frac{U_r+\bar U_r}2\right)^{-1/3}, \\
Z_{\mathcal I_{ab}^{1\bar1}} &=& 
\frac1{2^3} \left(\frac{S+\bar S}2\right)^{-1} 
\left(\frac{U_1+\bar U_1}2\right)^{-1/3}
\left(\frac{U_2+\bar U_2}2\right)^{-5/6}
\left(\frac{U_3+\bar U_3}2\right)^{-5/6}\\
&&\times\left(\prod_{k\neq 1}|M_{ab}^{(r)}|\right)^{-1/2}
{\rm exp}\left[-\sum_{k\neq1}\frac{4\pi}{U_k+\bar U_{\bar k}}
\frac{\left(\im\zeta_{ab}^{(k)}\right)^2}{M_{ab}^{(k)}}\right], \\
Z_{\mathcal I_{ab}^{j\bar j}} &=& 
\frac14 \left(\frac{T_j+\bar T_j}2\right)^{-1} 
\left(\prod_{r}\frac{U_{r}+\bar U_{r}}2\right)^{-1/3}\\
&&\times\left(\frac{|M_{ab}^{(j)}|}{|M_{ab}^{(s)}|_{j\neq s\neq1}}\right)^{-1/2}
{\rm exp}\left[-\sum_{k\neq1}\frac{4\pi}{U_k+\bar U_{\bar k}}
\frac{\left(\im\zeta_{ab}^{(k)}\right)^2}{M_{ab}^{(k)}}\right]. \\
\end{eqnarray*} 
The relevant Yukawa couplings  
\begin{eqnarray*}
\lambda_{mn} \phi_2^{an}\phi_1^n\phi_3^{an}\qquad({\rm in~}S_{mn}),\\
\lambda_{\mathcal I_{ab}} \phi_1^{\mathcal I_{ab}}\phi_2^{bn}\phi_3^{an}\qquad({\rm in~}S_{mn}),\\
\lambda_{\mathcal I_{ab}\mathcal I_{bc}\mathcal I_{ca}}^{ijk} 
\phi_i^{\mathcal I_{ab}}\phi_j^{\mathcal I_{bc}}\phi_k^{\mathcal I_{ca}}\qquad({\rm in~}S_m),
\end{eqnarray*} 
are given by 
\begin{eqnarray*}
\lambda_{mn} &=& 1,\\
\lambda_{\mathcal I_{ab}} &=& -1
\left(\prod_{r=2,3}
\vartheta\begin{bmatrix}
I_{ab}^{(r)}/M_{ab}^{(r)}\\0\end{bmatrix}\left(\zeta_{ab}^{(r)}, iM_{ab}^{(r)}\bar U_r\right)
\right)\\
\lambda_{\mathcal I_{ab}\mathcal I_{bc}\mathcal I_{ca}}^{ijk} 
&=& -\frac{1}3\epsilon^{\rm ijk} 
\delta_{\rm i}^{~i}\delta_{\rm j}^{~j}\delta_{\rm k}^{~k}
\prod_{r=2}^3 
\lambda^{(r)}_{I^{(r)}_{ab} I^{(r)}_{bc} I^{(r)}_{ca}},  
\end{eqnarray*} 
where $\lambda^{(r)}_{I^{(r)}_{ab} I^{(r)}_{bc} I^{(r)}_{ca}}$ is defined in 
Eqs.~ (\ref{eq:yukafinxx}) and (\ref{eq:finalyukaholo}). 
Note that $\phi_1^{\mathcal I_{ab}}$ and $\phi_{j\neq1}^{\mathcal I_{ab}}$ are 
clearly distinguished in the above expressions, because 
$\phi_{j\neq1}^{\mathcal I_{ab}}$ carries vector components of the 8D field theory but 
$\phi_1^{\mathcal I_{ab}}$ does not. 
In these expressions, the shift $\chi_s$ of the point-like localization of the 4D $U(N)$ SYM 
theory on the second and the third tori are absent, $\chi_s=0$. 
One can easily introduce the shift in the same manner as is in the last part of 
the previous section. 

We have derived the 4D effective supergravity action from the 4D and 8D SYM system in the 
$\mathcal N=1$ superfield description. 

\section{Conclusions and discussions}
\label{sec:cd}
A systematic way of dimensional reduction for 10D magnetized $U(N)$ SYM theories 
provided in Ref.~\cite{Abe:2012ya} has been extended, in this paper, to those for 
$(4+2n)$-dimensional $U(N)$ SYM theories ($n=0,1,2,3$) and their mixtures 
wrapping magnetized tori, which are described by 4D ${\mathcal N}=1$ superfields. 
Such a superfield description makes the $\mathcal N=1$ SUSY manifest, which is a 
(common) part of ${\mathcal N}=2,3$ or $4$ SUSY in the (mixture of) $(4+2n)$-dimensional 
SYM theories, preserved by the configuration of magnetic fluxes. While the magnetic 
fluxes break the higher-dimensional SUSY, the ${\mathcal N}=1$ SUSY is preserved as 
long as the auxiliary fields in ${\mathcal N}=1$ superfields have a vanishing VEV. 

It is important to study $\mathcal N=1$ SUSY configurations of magnetic fluxes 
from both phenomenological and theoretical points of view. It is known that 
non-SUSY configurations are generically unstable in string theory due to the 
appearance of tachyonic modes in various sectors. The ${\mathcal N}=1$ superfield 
description of higher-dimensional SYM theories~\cite{Marcus:1983wb,ArkaniHamed:2001tb} 
is so powerful to find the desired flux configurations and explicit forms of K\"ahler 
metrics and holomorphic Yukawa couplings with certain moduli dependences in the 
4D effective SUGRA. Since the moduli-mediated contributions to soft SUSY-breaking 
parameters are determined by them, it is easy to evaluate the induced SUSY spectra. 
This is a great advantage in phenomenological studies. 

Two concrete examples for mixed SYM systems wrapping magnetized tori have been shown. 
The first one consists of 6D $U(M)$ and 10D $U(N)$ SYM theories accompanied by their 
couplings given in Sec.~\ref{sec:6d10d}. This is a straightforward extension of the 
previous work~\cite{Abe:2012ya} based on a single 10D SYM theory and is derived from 
the 10D $U(M+N)$ SYM theory by introducing infinite numbers of magnetic fluxes, which 
is a useful tool to construct mixed SYM systems in a systematic way. Especially, the 
bifundamental representations crossing over the two SYM theories, $(M, \bar N)$ and 
$(\bar M, N)$, those can couple to both 6D $U(M)$ and 10D $U(N)$ adjoint fields, 
will be strongly localized in the vicinity of the 6D hypersurface in which the 
6D SYM fields reside if they are identified as open-string modes in D5/D9 systems. 

In the field theoretical description, the infinite magnetic fluxes induce such a 
point-like localization~\cite{Cremades:2004wa}, and the well-defined delta functions 
are obtained as solutions of zero-mode equations for the bifundamental representations. 
Such a procedure utilizing infinite fluxes to construct a mixed SYM system is motivated 
by a T-duality in D-brane systems. It is known that D9-branes with infinite magnetic 
fluxes in four compact directions are related to pure D5-branes without magnetic fluxes 
by the T-duality in these directions. 

At the last step to derive 4D effective SUGRA, we promote gauge coupling constants 
and torus parameters to moduli fields in accordance with modified parameterizations 
of the moduli VEVs. The modifications are required to describe moduli in a universal 
way in mixed SYM theories with different dimensionalities from each other, because 
their gauge coupling constants have different mass dimensions depending on their 
dimensionalities. The modified parameterizations could also have been interpreted 
consistently in a D-brane picture. 

Another example which consists of 4D $U(N)$ and 8D $U(M)$ SYM theories have been 
shown explicitly in Section~\ref{sec:4d8d}. This SYM system is also derived from 
the 10D $U(M+N)$ SYM theory with the same configuration of infinite magnetic fluxes 
as that of the previous example. This seems plausible because D3/D7 brane systems 
are T-dual to D9/D5 brane systems in the framework of type IIB compactifications. 
Referring to the duality in such D-brane pictures, we have adopted another 
parameterization for the VEVs of moduli fields in this second example. 
We confirmed a validity of the moduli parameterization in each example by comparing 
the obtained moduli-dependences of gauge kinetic functions in the 4D effective SUGRA 
with those identified in the corresponding D-brane system. They exactly accord with 
the corresponding ones in the D-brane picture. 

Although we have shown only two examples, a wide variety of combinations of multiple 
SYM theories can be realized in the same manner. Such a variety is expected to be of 
service in phenomenological/cosmological studies towards a realistic model. 
For instance, these multiple-SYM systems would provide a foundation for constructing 
moduli stabilization and dynamical SUSY breaking sectors, which are 
desired to be sequestered from the visible sector from the phenomenological point 
of view. In such a construction, bifundamental fields charged under both hidden and 
visible sectors could appear depending on the flux configuration, some of those 
can play a role of messenger which mediates SUSY breaking contributions from the 
hidden to the visible sector. Such a gauge-mediated contribution~\cite{Dine:1981gu} 
to soft SUSY-breaking parameters can be one of the distinctive features of the system. 

On the other hand, phenomenological consequences of mixed moduli- and anomaly-mediated 
contributions to the soft parameters (that is called mirage mediation~\cite{Endo:2005uy}) 
were studied in a model based on the 10D magnetized SYM theory~\cite{Abe:2012fj,
Abe:2014soa}. If the model is extended to a mixed SYM system where the dynamical SUSY 
breaking sector is incorporated, the gauge-mediated contribution can also be comparable 
to moduli- and anomaly-mediated ones with a certain moduli stabilization mechanism. 
In such a case the system provides a UV completion of the deflected mirage 
mediation~\cite{Everett:2008qy}. In any case, the low-energy spectra in visible 
and hidden sectors are governed by the configuration of background magnetic fluxes 
in the SYM system. 

The gauge kinetic function in the 4D effective action is given by the dilaton $S$ or 
the K\"ahler modulus $T_i$, depending on the dimensionality of the original SYM theory. 
If the SM gauge groups originate from different SYM theories in the mixed system, 
certain non-universal gauge kinetic functions can be realized in the SM sector. 
In such a case, some attractive scenarios are then conceivable deviating from the 
grand unification models, especially, non-universal gaugino masses at the 
compactification scale are possible at the tree-level, even when the gauge coupling 
constants are unified at the same scale. It is known that a certain value of 
wino-to-gluino mass ratio extremely relaxes a fine-tuning of Higgsino-mass parameter 
(so-called $\mu$-parameter) required for triggering a correct electroweak symmetry 
breaking in the MSSM or MSSM-like models~\cite{Abe:2007kf} without conflicting with 
the observed Higgs boson mass at the Large Hadron Collider~\cite{Abe:2012xm}. 

Furthermore, in the moduli stabilization and SUSY breaking sectors, the moduli 
dependences of their gauge kinetic functions are extremely important, because 
some nonperturbative effects induced by the SYM dynamics are usually required 
in these sectors. For example, in the KKLT scenario of moduli stabilization~\cite{
Kachru:2003aw}, K\"ahler-moduli dependent nonperturbative effects are assumed 
which determine the ratio between moduli- and anomaly-mediated SUSY breaking~\cite{
Choi:2004sx}. We should remark that there appear stringy corrections which mix 
multiple moduli in each gauge kinetic function depending on the configuration of 
magnetic fluxes~\cite{Lust:2004cx,Blumenhagen:2006ci}, when the SYM system is 
treated as a low-energy effective description of D-branes. Such a moduli-mixing 
in the gauge kinetic functions plays a role in the mechanism of moduli stabilization 
and SUSY breaking~\cite{Abe:2005rx}. 

While these SYM theories in various-dimensional spacetime are related to each other 
by the T-duality in a D-brane picture, there are differences in their K\"ahler 
metrics and holomorphic Yukawa couplings, because their moduli dependence depends 
on the configurations of (finite) magnetic fluxes. The dynamics of moduli fields 
in low-energy effective field theories is quite significant in particle physics and 
cosmology, especially, in the early universe. This has recently attracted much 
attentions as cosmological observations highly evolve. In the study of early universe, 
couplings between the moduli and the matter particles have to be treated carefully. 
Since the higher-dimensional SYM systems give explicit forms of the couplings, 
it is of great interest to study these systems incorporating a certain scenario 
of the early universe. 

The D-brane pictures, especially T-dualities in type II superstring theories, 
motivate and support this work. Indeed, in this paper, we find them in many respects. 
Although there are several issues to be addressed, such as tadpole cancellations~\cite{
Angelantonj:2000hi,Blumenhagen:2006ci}, for a string realization of the mixed SYM 
system treated here, it is worth trying and we will study further elsewhere.

\subsection*{Acknowledgement}
H.A. was supported in part by the Grant-in-Aid for Scientific Research No.~25800158 
from the Ministry of Education, Culture, Sports, Science and Technology (MEXT) in Japan. 
K.S. was supported in part by a Grant-in-Aid for JSPS Fellows 
No.~25$\cdot$4968 from the MEXT in Japan. 

\appendix

\section{A SUSY configuration in 6D and 10D SYM systems}
\label{app:susy6d10d}
We have adopted the generic configuration of magnetic fluxes 
in the 6D and 10D SYM theories in Subsection~3.2 for completeness of our description, 
even though it would break the $\mathcal N=1$ SUSY. 
That is, the magnetic flux in 6D $U(M)$ sector $M_m^{(1)}$ is nonvanishing 
on the first torus on which the 6D SYM theory is compactified, 
and then, 
the gauge symmetry is broken by it as $U(M)\rightarrow \prod_aU(M_a)$. 
In this appendix, we calculate 4D effective SUGRA on the basis of another configuration with 
vanishing $M_m^{(1)}$, where the $U(M)$ gauge group is preserved as well as 
the $\mathcal N=1$ SUSY. 
There are four adjoint fields in the superfield description of 6D $U(M)$ theory. 
We denote their zero-modes as 
\begin{equation*}
V^{m,n_1=0}\equiv V^{m},\qquad 
\phi_i^{m,n_1=0}\equiv g_6 \phi_i^m,  
\end{equation*}
where chiral superfields are normalized by the gauge coupling $g_6$ for convenience. 
Since these $U(M)$ adjoint fields do not feel magnetic fluxes on the first torus, 
their extra-dimensional wavefunctions are flat. 
The integration with respect to the first torus coordinates $z_1$ and $\bar z_{\bar1}$ 
can be straightforwardly performed, 
and it is easy to derive the 4D effective action. 
According to the normalization (\ref{eq:normal}), 
their K\"ahler metric $\tilde Z^{\bar i j}$ 
and tri-linear coupling $\tilde\lambda^{ijk}$ 
are given by 
\begin{eqnarray}
\tilde Z^{\bar i j} &=& 2h^{\bar ij}\nonumber\\
\tilde\lambda^{ijk} &=& 
-\frac{2g_6}3\epsilon^{\rm ijk}e_{\rm i}^{~i}e_{\rm j}^{~j}e_{\rm k}^{~k}  
(\mathcal A^{(r)})^{-1/2}, \label{eq:appenzl}
\end{eqnarray}
instead of Eqs.~(\ref{eq:6dtildezz}) and (\ref{eq:6dtildell}).  

We can find the 4D effective SUGRA on this background in the same manner 
as is in Sec.~3. 
The rescaling rules are given by 
\begin{eqnarray*}
\phi_i^m&\rightarrow&\alpha_i^m\phi_i^m\qquad{\rm for~}i=1,2,3,
\end{eqnarray*}
where 
\begin{eqnarray*}
\alpha_1^m &=& 2^{-1} e^{-\langle\phi_4\rangle} 
\frac{2\pi R_1}{\sqrt{\re \langle S\rangle }}, \\
\alpha_j^m &=& 2^{-7/4} e^{-\langle\phi_4\rangle} 
\frac{2\pi R_j}{\sqrt{\re \langle T_j\rangle }} 
\left(\prod_{r} \re \langle U_{r}\rangle \right)^{-1/4}  \qquad{\rm for}~j=2,3. \\
\end{eqnarray*} 
After these rescalings, the parameters are 
promoted to the dilaton and moduli superfields in the K\"ahler potential 
in accordance with Eqs.~(\ref{eq:gaugeconst}) and (\ref{eq:modulidef59}). 
We find 
\begin{eqnarray*}
Z^{\bar 11} &=& 
\frac1{2} \left(\frac{S+\bar S}2\right)^{-1} , \\
Z^{\bar jj} &=& 
\frac1{2^{5/2}} \left(\frac{T_j+\bar T_j}2\right)^{-1} 
\left(\prod_{r} \frac{U_r+\bar U_r}2 \right)^{-1/2} 
\qquad {\rm for~}\bar j=j=2,3, \\
\end{eqnarray*} 
and then, the tri-linear coupling $\tilde\lambda^{ijk}$ is simply given by 
\begin{eqnarray*}
\lambda^{ijk} &=& 
-\frac13\epsilon^{\rm ijk}e_{\rm i}^{~i}e_{\rm j}^{~j}e_{\rm k}^{~k} .  
\end{eqnarray*}

We should remark on couplings between these adjoint fields and 
$(mn)$-fields. 
There are bifundamental representations charged under the $U(M)$ gauge group 
and a $U(N_a)$ gauge subgroup in this system. 
If a pair of representations $(M, \bar N_a)$ and $(\bar M, N_a)$ appears 
in the 4D effective theory, 
they can couple to the above adjoint representations. 
However, either of these two bifundamental representations 
is eliminated by the chirality projection due to magnetic fluxes 
because these two representations feel magnetic fluxes with opposite signs 
(The bifundamentals are contained in only $\phi_2$ and $\phi_3$, 
which require the negative sign of magnetic fluxes on the first torus 
for their zero-modes to survive ). 
As a result, the $U(M)$ adjoint fields will not couple to the other sectors 
unless representations $(M, \bar N_a)$ and $(\bar M, N_a)$ feel vanishing fluxes.

\end{document}